\documentclass[manuscript,screen]{acmart}

\usepackage{algorithmic}
\usepackage{graphicx}
\usepackage{flushend}
\usepackage{subcaption}
\usepackage{adjustbox}

\newcommand{\ignore}[1]{}

\newcommand{\blueHL}[1]{{\textcolor{black}{#1}}}
\newcommand{\greenHL}[1]{\textcolor{black}{#1}}

\usepackage{multirow}
\usepackage{textcomp}
\usepackage{xcolor}
\usepackage{enumitem}

\setcopyright{acmlicensed}
\acmJournal{TODAES}
\acmYear{2023} \acmVolume{1} \acmNumber{1} \acmArticle{1} \acmMonth{9} \acmPrice{15.00}\acmDOI{10.1145/3526115}

\begin{document}

\title{Bridging the Gap between Global Route and Detailed Route \\ Using ML for Wire Parasitics and Delay Prediction}
\title{Enhancing Timing Closure with ML-assisted Parasitics and Delay Prediction for Post-Global Route Optimization}
\title{A Machine Learning Approach to Improving Timing Consistency between Global Route and Detailed Route}

\renewcommand{\shorttitle}{}
% Apparently, author names cannot be grouped together, but affiliations can.

\author{Vidya A. Chhabria} 
\affiliation{%
  \institution{Arizona State University}
  \city{Tempe}
  \state{AZ}
  \country{USA}
}
\author{Wenjing Jiang}
\authornote{Primary author}
\affiliation{%
  \institution{University of Minnesota}
  \city{Minneapolis}
  \state{MN}
  \country{USA}
}

\author{Andrew B. Kahng}
\affiliation{%
  \institution{University of California -- San Diego}
  \city{La Jolla}
  \state{CA}
  \country{USA}
}

\author{Sachin S. Sapatnekar}
\affiliation{%
  \institution{University of Minnesota}
  \city{Minneapolis}
  \state{MN}
  \country{USA}
}

\renewcommand{\shortauthors}{Vidya A. Chhabria, Wenjing Jiang, Andrew B. Kahng, \& Sachin S. Sapatnekar}

\begin{abstract}
Due to the unavailability of routing information in design stages 
prior to detailed routing (DR), the tasks of timing prediction and optimization pose major challenges.
Inaccurate timing prediction wastes design effort, hurts circuit performance, and may lead to design failure.  
This work focuses on timing prediction after clock tree synthesis and 
placement legalization, which is the earliest opportunity to time and optimize a
``complete'' netlist. The paper first documents that having ``oracle knowledge'' of the final post-DR
parasitics enables post-global routing (GR) optimization to produce improved final timing outcomes. 
To bridge the gap between GR-based parasitic and timing estimation and post-DR results {\em during post-GR optimization},
machine learning (ML)-based models are proposed, including the use of features for macro blockages 
for accurate predictions for designs with macros. Based on a set of experimental evaluations, it is
demonstrated that these models show higher accuracy than GR-based timing estimation.  When 
used during post-GR optimization, the ML-based models show demonstrable improvements in post-DR circuit 
performance. The methodology is applied to two different tool flows -- OpenROAD and a commercial tool flow --
and results on an open-source 45nm bulk and a commercial 12nm FinFET enablement show improvements in post-DR timing slack 
metrics without increasing congestion.  The models are demonstrated to be generalizable to designs 
generated under different clock period constraints and are robust to training data with small levels of noise.
\end{abstract}




\maketitle

\section{Introduction}
\label{sec:Introduction}

Interconnect effects are a significant factor in determining the post-layout performance 
of digital designs in modern integrated circuit technology nodes.
In particular, due to increased per-unit length resistances and capacitances from one
technology generation to the next, wire delays have become a significant bottleneck in
achieving design closure and overall IC performance outcomes.  The accurate estimation and
prediction of wire parasitics and delays is particularly important in modern design flows, as
incorrect estimates can result in unnecessary design iterations; in some cases, it may
be impossible to achieve design closure on schedule, which may result in increased time to market.  If wire 
parasitics and delays can be estimated well, they can be used to guide timing optimizations 
such as net buffering and logic gate resizing at multiple stages of the RTL-to-GDS
implementation flow.

\begin{figure}[tbp]
\centerline{\includegraphics[width=0.6\linewidth]{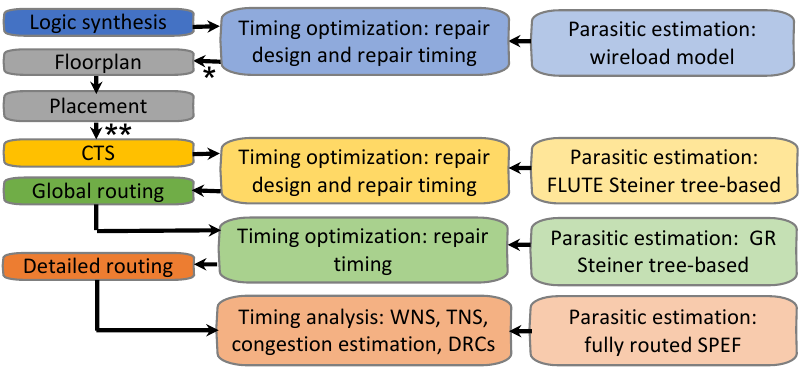}}

\caption{Timing optimizations in the physical design flow with different parasitic 
estimates. In modern production flows, the output of vendor A's synthesis
tool is ``de-buffered'' when passed to vendor B's place-and-route (P\&R) tool. Then,
ERC compliance is enforced during global placement with buffering and resizing.
The de-buffering and ERC-fixing steps are respectively marked with 
(*) and (**) in the figure.}
\label{fig:traditional-flow}
\end{figure}

Fig.~\ref{fig:traditional-flow} shows a standard physical design flow that 
highlights multiple timing optimization steps.  It is essential to perform 
optimization several times in the flow so that netlist changes between successive
stages are manageable within a convergent design methodology. This is because a
too-drastic netlist change between flow stages can force looping back to earlier 
steps of the flow, rather than continuing forward. However, due to the unavailability of routing 
information before the final stages, perfectly accurate wire delay estimates are
impossible.

To overcome this challenge, design flows use different models to account for wire 
parasitics during timing optimizations based on the information available at each 
stage. For example, as highlighted in Fig.~\ref{fig:traditional-flow}, wireload 
models are used for gate-level optimization during logic synthesis. During global
placement and even after clock tree synthesis (CTS), generic half-perimeter wirelength (HPWL) or FLUTE-based~\cite{Chu08} Steiner 
tree estimates, scaled by layer-averaged per-unit resistances and capacitances, are used 
for electrical rule check (ERC, including max load and max transition rules) compliance and 
some gate-level optimizations.  However, these 
models can be highly inaccurate (overly conservative or grossly optimistic) compared 
to the true parasitics and timing estimates after detailed route. 
Any optimization or synthesis using these models can produce 
solutions that are either overdesigned (pessimistic) or underdesigned (optimistic, i.e., failing
electrical and/or performance constraints). 

As a design proceeds from early stages (e.g., RTL specification, floorplanning) to later stages 
(e.g., detailed placement, detailed routing), an increasing amount of physical information (i.e., spatial embedding) is
available, leading to potentially better estimates of parasitics. However, even at relatively
late stages in the physical design flow, such as routing, there can be significant estimation
inaccuracy.  As shown in Fig.~\ref{fig:traditional-flow}, a design is typically routed in two stages: 
global routing (GR) and detailed routing (DR).  
The GR step allocates routing resources to each net, and generates a routing plan 
that the DR step takes as initial guidance toward a final routing solution. 
In modern tools, the routing plan from GR is represented in the form of {\em route guides} that 
contain  information on layer assignments and Steiner tree topologies for each 
net~\cite{DolgovVWX19}\cite{Kahng-TritonRoute21}, and these are used to guide DR. Clearly,
the ability to close timing depends on the quality of the GR route guides.

Importantly, the GR stage is typically followed by a timing optimization step 
(sizing, fanout clustering, buffering, etc.) as well as a final (post-optimization)
placement legalization step, since better parasitic estimates are available post-GR
compared to after earlier flow stages. 
However, GR-based parasitic estimates are still inaccurate relative to final 
DR outcomes, as they do not fully comprehend such factors as detailed design 
rules, pin access challenges, and congestion, which are glossed over in GR. 
Together, these factors cause wire detours and layer
reassignments during DR, which in turn cause GR-based estimates and 
DR-based outcomes to diverge.

\begin{figure}[tb]
\centerline{\includegraphics[width=0.98\linewidth]{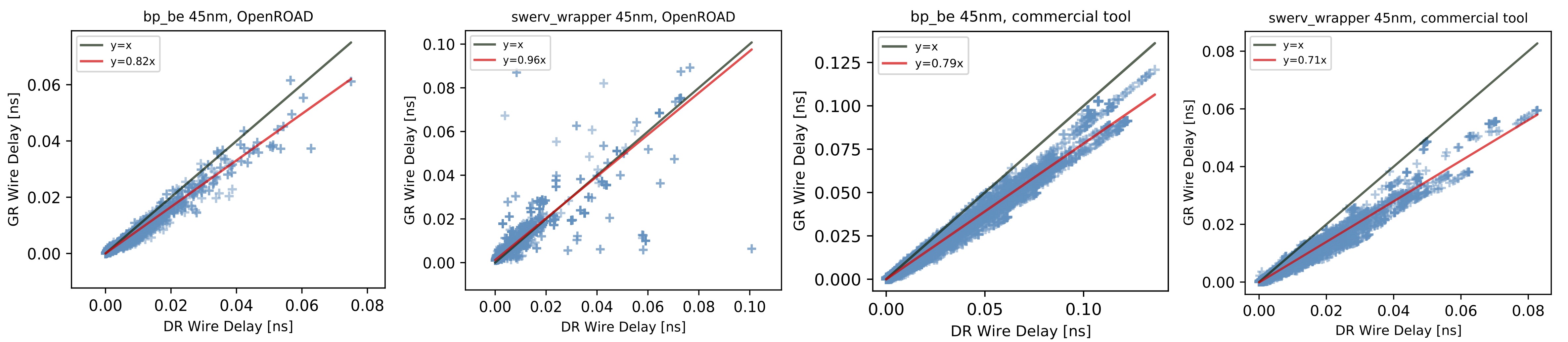}}
\caption{\blueHL{Discrepancy between post-GR and post-DR wire delays of bp\_be 45nm and swerv\_wrapper 45nm in OpenROAD (left figures) and 
\greenHL{bp\_be 45nm and swerv\_wrapper 45nm in a commercial tool flow (right figures)}. The slope $k$ and the intercept $b$ in the best-fit $y = kx+b$ (red traces) differ across tool flows and across designs.}}
\label{fig:linear-fit}
\end{figure}

Fig.~\ref{fig:linear-fit} shows the inaccuracy in wire delay 
estimation when using a GR-based model that estimates wire parasitics using 
FLUTE-generated~\cite{Chu08} Steiner trees from FastRoute~4.1~\cite{Pan12},
as compared to the wire delay of corresponding detail-routed nets when using 
RC trees from post-DR parasitic extraction. The figures show discrepancies in 
\blueHL{the estimated wire delay of \greenHL{four designs: bp\_be (42K nets) and swerv\_wrapper (88K nets),
each implemented in 45nm 
technology in both OpenROAD and commercial tool flows.} Each plot shows a $y=x$ line: a
perfect prediction would lie along this line; an optimistic prediction, where the 
GR-based delay estimate is smaller than the DR-based ground-truth delay, would lie below
the line; and a pessimistic prediction would lie above the line.
In OpenROAD, the GR-stage estimates are largely optimistic 
for bp\_be, but may be either pessimistic or optimistic for swerv\_wrapper. 
For the commercial tool flow, the GR-based wire delay estimates for the most part are optimistic.
While it could be argued that the prediction can be corrected by using a
best linear fit line ($y = kx+b$) instead of the $y=x$ line, this does not repair
all of the discrepancies.  Specifically, we draw the following conclusions:
% \redfn{Why do you say $y=kx+b$? Both plots seem to be of the form $y=kx$ with $b=0$ (according to the legends).~\blueHL{$y=kx+b$ is used for linear fit, but $b$ value is very small}} 
\begin{itemize}
    \item The slope $k$ for calibrating the best-fit curve is tool-flow-specific, and we see different
    values for the OpenROAD flow and the commercial tool flow.
    \item Even for the same tool flow, the best-fit curve is design-specific: the value
    of $k$ for bp\_be is significantly different from that for swerv\_wrapper. \greenHL{This
    is observed to be true for both the OpenROAD and commercial tool flows.}
    \item Compared to the discrepancy observed in OpenROAD, the wire delay discrepancy 
    for the commercial tool flow is lower, reflecting a superior ability to adhere to GR
    route guides.  Even so, the discrepancy is significant.
\end{itemize}
Therefore, the problem of predicting post-DR interconnect delays from post-GR route guides
is more complex than a simple correction through a curve-fit\greenHL{, and this motivates
our approach of using an ML predictor.}}

\begin{table}[tb]
\caption{\blueHL{Comparison of post-DR WS when using GR-based vs. DR-based
parasitics for post-GR timing optimizations.}}
\label{tbl:post-DR-motivation}
\resizebox{0.6\linewidth}{!}{%
\begin{tabular}{|c|c|c|c|c|c|c|c|} 
\hline
\multirow{2}{*}{\textbf{Design}} & \multirow{2}{*}{\begin{tabular}[c]{@{}c@{}}\textbf{CLK}\\\textbf{(ns)}\end{tabular}} & \multirow{2}{*}{\textbf{Tech}} & \multirow{2}{*}{\textbf{\#Nets}} & \multirow{2}{*}{\textbf{\#Macros}} & \multirow{2}{*}{\greenHL{\textbf{Utilization}}} & \multicolumn{2}{c|}{\textbf{Post-DR WS (ns)}}                                                                                \\ 
\cline{7-8}
                                 &                                                                                      &                                &                                  &                                    &                                       & \begin{tabular}[c]{@{}c@{}}GR-based\\parasitics\end{tabular} & \begin{tabular}[c]{@{}c@{}}DR-based\\parasitics\end{tabular}  \\ 
\hline
riscv32i                         & 9.6                                                                                  & \multirow{4}{*}{130nm}         & 8150                             & 0                                  & \greenHL{67.52\%}                               & -0.26                                                        & -0.26                                                         \\ 
\cline{1-2}\cline{4-8}
aes                              & 5.4                                                                                  &                                & 15307                            & 0                                  & \greenHL{62.91\%}                               & -0.21                                                        & -0.19                                                         \\ 
\cline{1-2}\cline{4-8}
ibex                             & 16.0                                                                                 &                                & 15369                            & 0                                  & \greenHL{55.85\%}                               & -0.56                                                        & -0.60                                                         \\ 
\cline{1-2}\cline{4-8}
jpeg                             & 7.8                                                                                  &                                & 59573                            & 0                                  & \greenHL{62.90\%}                               & -0.25                                                        & -0.17                                                         \\ 
\hline
dynamic\_node                    & 1.0                                                                                  & \multirow{7}{*}{45nm}          & 11598                            & 0                                  & \greenHL{57.63\%}                               & -0.25                                                        & -0.17                                                         \\ 
\cline{1-2}\cline{4-8}
ibex                             & 2.0                                                                                  &                                & 16836                            & 0                                  & \greenHL{58.94\%}                               & -0.24                                                        & -0.11                                                         \\ 
\cline{1-2}\cline{4-8}
aes                              & 0.8                                                                                  &                                & 17566                            & 0                                  & \greenHL{59.60\%}                               & -0.27                                                        & -0.09                                                         \\ 
\cline{1-2}\cline{4-8}
jpeg                             & 1.4                                                                                  &                                & 68247                            & 0                                  & \greenHL{52.48\%}                               & 0.24                                                         & 0.02                                                          \\ 
\cline{1-2}\cline{4-8}
swerv\_wrapper                   & 2.5                                                                                  &                                & 88490                            & 28                                 & \greenHL{45.87\%}                               & -0.24                                                        & -0.23                                                         \\ 
\cline{1-2}\cline{4-8}
bp\_fe                           & 2.2                                                                                  &                                & 24883                            & 11                                 & \greenHL{43.73\%}                               & -0.15                                                        & -0.10                                                         \\ 
\cline{1-2}\cline{4-8}
bp\_be                           & 2.8                                                                                  &                                & 41973                            & 10                                 & \greenHL{38.30\%}                               & -0.15                                                        & -0.12                                                         \\ 
\hline
swerv\_wrapper                   & 1.2                                                                                  & \multirow{2}{*}{12nm}          & 92787                            & 28                                 & \greenHL{49.08\%}                               & -0.48                                                        & -0.24                                                         \\ 
\cline{1-2}\cline{4-8}
coyote                           & 3.2                                                                                  &                                & 272948                           & 15                                 & \greenHL{49.00\%}                               & -0.27                                                        & -0.15                                                         \\
\hline
\end{tabular}
}
\end{table}
\smallskip
\noindent
{\bf ``Oracle'' knowledge of post-DR parasitics improves final outcomes.}
As explained earlier, the use of inaccurate parasitic and timing estimates in
any timing optimization can lead to harmful pessimism (overdesign that wastes
resources) or optimism (underdesign that leads to design iterations). 
%pessimistic or optimistic in suboptimal (inefficient buffering and resizing)  solutions that hurt circuit performance (post-DR circuit delay), and also create  inconsistencies between GR and DR which may increase the number of iterations  between GR and DR  during design closure. 
We have conducted a motivating study to show the potential benefit of ``oracle''
knowledge of post-DR wire parasitics, were these parasitics to somehow be 
available to post-GR timing optimizations. 
\blueHL{Table~\ref{tbl:post-DR-motivation} shows results on four open-source 130nm~\cite{SkyWater130} testcases, seven open-source 
45nm~\cite{NanGate45} testcases and two commercial 12nm 
testcases, using an open-source flow~\cite{ORFS}. As indicated in the table, some of these layouts consist of 
standard cells only, with no macros,
while others contain a mix of standard cells and large macro blocks. The table
highlights the cost of post-GR buffering and resizing solutions 
that are driven by inaccurate parasitics, \greenHL{and also lists the utilization for each
design}.  For example, if the discrepancy 
as in Fig.~\ref{fig:linear-fit} is corrected in post-GR for each design, the improvement 
in the post-DR worst slack (WS) for these designs can be as much as
240ps ($-0.48$ns $\rightarrow$ $-0.24$ns).}
We ascribe the WS improvement to the 
early identification of true (post-DR) 
timing violations, which allows post-GR optimizations to efficiently buffer nets 
and resize logic gates on truly critical timing paths. In the absence of accurate
prediction, these timing paths might be 
missed due to optimism, or unnecessarily buffered and resized due to pessimism. 
This motivates the key result of our work, which is to apply machine learning (ML) to 
close the post-GR-to-post-DR parasitic estimation gap. 
\greenHL{On closer inspection, we see that the designs with smaller post-DR WS improvement 
in Table~\ref{tbl:post-DR-motivation} correspond to high values of utilization. For example, 
riscv32i has the highest utilization among 130nm designs without macros, and swerv\_wrapper 
has the highest utilization among 45nm designs with macros. In these designs, 
\greenHL{fewer options are available for resizing and buffering in post-GR timing optimization, 
resulting in fewer perturbations due to these operations. This may partially explain why post-DR WNS from the flows 
based on GR-based and DR-based parasitics are quite close.}}

\begin{figure}[tb]
\centerline{\includegraphics[width=0.6\linewidth]{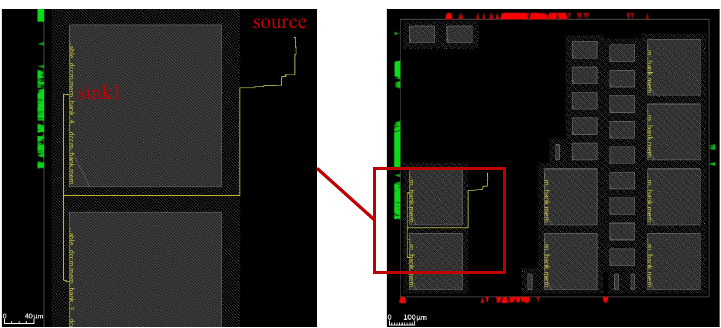}}
\bigskip
\begin{tabular}{|l|l|l|} 
\hline
                             & Post-GR & Post-DR  \\ 
\hline
Total Capacitance (pF)     & 0.0685  & 0.0893   \\ 
\hline
Total Resistance (k$\Omega$) & 1.0468  & 1.7973   \\ 
\hline
Wire delay (ns)              & 0.0354  & 0.1059   \\
\hline
\end{tabular}
\caption{\blueHL{Net detours due to macro blockage in swerv\_wrapper, 45nm. Total capacitance, total resistance and wire delay of source-sink1 in the sample net.}} 
\label{fig:net_detour_macro_OR}
\end{figure}

\smallskip
\blueHL{While significant inaccuracies are seen even for designs without macros, the presence of large macros in a design is a source of another major degree of difficulty in
delay prediction. Depending on their structure, these macros may act as partial or complete
blockages for interconnect wires, depending on whether some or all metal layers are blocked by the macro.  The 
wiring detours that are required to circumvent the blockages can
greatly impact the routing solution -- notably, the wirelength, wire delay, and the need for buffer insertion
along these wires.  As an example, Fig.~\ref{fig:net_detour_macro_OR} shows a placement of swerv\_wrapper in 45nm 
with several macros. The figure at right shows the overall layout, with a region of interest shown within a red
rectangle, with two large macros placed close to each other.  At left, we zoom into this rectangle, highlighting 
a three-pin net that connects a source pin at one side of two macros to two sinks at the other side of the macros: 
clearly this routing solution is required to detour around these macro blockages. From source to sink1, the wire bypasses 
a macro of size 206.9$\mu$m$\times$219.8$\mu$m through the halo between two macros, and the source-sink1 wire 
is 30\% longer than the Manhattan distance between the two pins. The impact of the detour on key performance metrics
for the net are listed in the table at the bottom of the figure: its post-DR capacitance and resistance are,
respectively, 22.5\% and 71.7\% higher than those predicted at the end of GR, and its wire delay discrepancy 
between the post-GR and post-DR steps is 199.2\%. Therefore, a critical characteristic of any 
prediction model must be its capability to predict interconnect delays for designs with macros.}

\begin{figure}[tb]
\centerline{\includegraphics[width=0.5\linewidth]{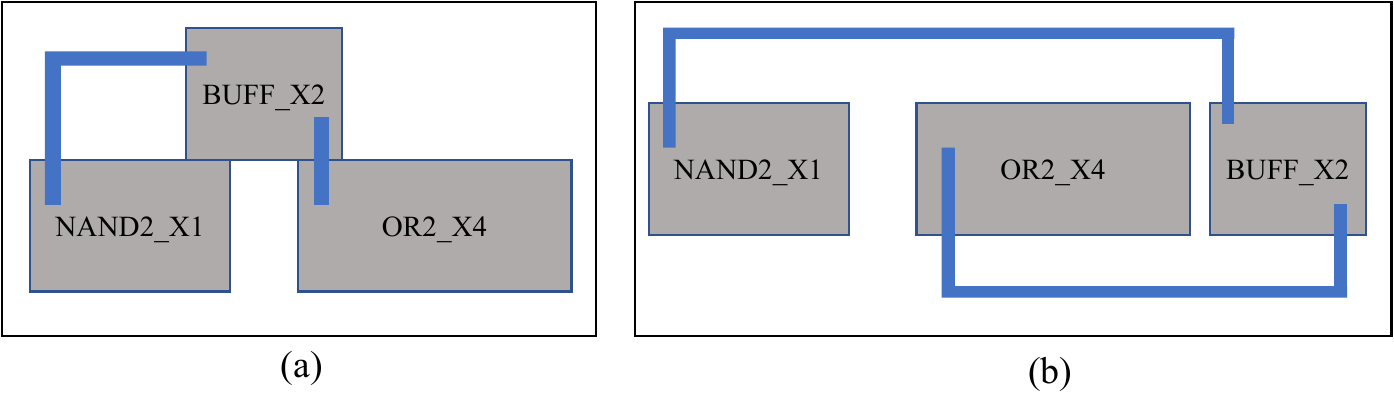}}
\caption{\blueHL{The effect of different buffer locations on the routing results: (a) The inserted buffer is placed in a different row of the NAND gate and OR gate. (b) The inserted buffer is placed in the same row as the NAND gate and OR gate.}}
\label{fig:route_noise}
\end{figure}

\blueHL{Another practical consideration is related to the fact that EDA tools are inherently noisy \cite{Chan20}, and can provide different results for nearly-identical inputs and runscripts.  Therefore, the noisy data collected from these tools can affect the estimation of final results and may lead to incorrect buffering and sizing during optimization, thus affecting design quality. For example, a small perturbation in the location of an instance can result in a different route. This is illustrated in Fig.~\ref{fig:route_noise}, where a buffer is inserted between a NAND gate and an OR gate during timing optimization to reduce the delay of the NAND gate. However, depending on the precise location of the inserted buffer -- two possibilities are shown in Fig.~\ref{fig:route_noise}(a) and Fig.~\ref{fig:route_noise}(b) -- the delay between the NAND gate and OR gate may be different due to dissimilarities in the wire delay. 
% Such differences can greatly impact the delay estimation between NAND gate and OR gate before performing timing optimization. 
Therefore, the capability of estimation models to handle noise in the training data is another critical evaluation criterion. Due to the high cost of generating training data, it is not practical to explore the entire ``noise space'' around a particular training point; instead, we verify that the ML-based models we build are resilient to such noise.}

% \redHL{Wenjing: Could you add a paragraph here talking about noise and its importance? This will motivate the new
% results on noise and help justify the 30\% addition. If you can show a graphic to justify the origin of this noise,
% it would be very helpful. The tone should be similar to the above paragraph, where you use an example to justify why
% the ML model must consider macros.  Possibly, this could be related to the robustness of the model (e.g., in the sense
% of the gibbon-panda example in Fig.~1 of \url{https://arxiv.org/pdf/1412.6572}.}

\smallskip
\noindent
{\bf Related works.}  Several researchers have worked in the general area of ML-based delay prediction 
during physical design. 
For a specified net topology,~\cite{Cheng20} 
builds an XGBoost-based ML model for the wire delay of a net of fixed topology, 
trained on commercial parasitic extraction and timing analysis tools. The  
impact of macro block layout on timing is predicted using boosting and 
SVM in~\cite{Chan16}. The work in~\cite{Barboza19} solves the problem 
of predicting wire delay and slew based on placement results, prior to GR, while~\cite{Yang22} predicts path delays prior to
routing, based on placement features, using a transformer network and residual 
model. The work in~\cite{He22} uses a look-ahead RC network generated by a coarse routing step (decomposition of multi-pin nets into two-pin nets, then routed using L-shaped routes) on the placed design for feature extraction, and uses this to perform post-placement net-based timing prediction.  In~\cite{Guo22}, a GNN model is used to estimate prerouting slacks at the endpoints of the design. \blueHL{In~\cite{Ye23}, wire delay and wire slew are  predicted by using a graph learning architecture that encodes the information of the RC tree of a net.} 
\smallskip

\noindent
{\bf Contributions.}  
\blueHL{In this paper, we propose a set of techniques to enhance the accuracy of timing prediction on designs with macros in the post-GR stage to improve the quality of the DR routing solution. The key contributions of this work are as follows:}
\begin{enumerate}[leftmargin=5mm]
     \item We show how ML enables the fast and accurate prediction of post-DR parasitics and timing estimates using post-GR information,  \blueHL{particularly for design with macros, in both bulk and FinFET technology nodes}. 
    \item We apply the ML model to the OpenROAD~\cite{ajayi19} physical design flow and to a commercial flow, and show up to  \blueHL{0.24ns savings (12nm node) in post-DR worst slack (WS) for OpenROAD, and 0.32ns savings (45nm node) for the commercial tool flow, without degrading congestion. }    
    \item \blueHL{We demonstrate that a similar flow that uses ML models for timing estimation can also be applied within a commercial tool flow, operating within the constraints of the information available through the available APIs, to improve the timing prediction accuracy and DR solution quality.}
    \item We find that as compared to a traditional flow, with a small increase in runtime (to perform ML inference), the ML model can improve the mean \%error of path slack from 5.75\% to 1.15\% in a 45nm testcase, and from 14.91\% to 7.61\% in a 12nm testcase.
    \item \blueHL{The proposed timing prediction models are assessed to be generalizable with respect to designs generated with different clock constraints, and can handle small noise in datasets.}
\end{enumerate}
\blueHL{A preliminary version of this research was published in~\cite{Chhabria22}.  This work expands upon the prior version by incorporating the impact of macros in the layout, and by extending the methodology to multiple tool flows, testing our flows on commercial technology, and evaluating multiple aspects of both the open-source and commercial flows.}
\section{Preliminaries}
\label{sec:background}

\subsection{Post-GR and post-DR routing estimates}

Timing-driven physical design requires an estimate of the delay of each stage of logic.  This corresponds to the sum of the gate delay and the wire delay, each of which is dependent on wiring parasitics.  To predict circuit timing, post-GR interconnect parasitics may be estimated based on the route guides. The estimated RCs for a net are determined by the length and the topology of the GR-constructed Steiner tree used for GR and the layer assignments. In this work, we use FastRoute~\cite{Pan12}, which performs precise layer assignment for each route, i.e., the route guides specify the precise layer for each segment of the Steiner tree.

\begin{figure}[tb]
\centerline{\includegraphics[width=0.99\linewidth]{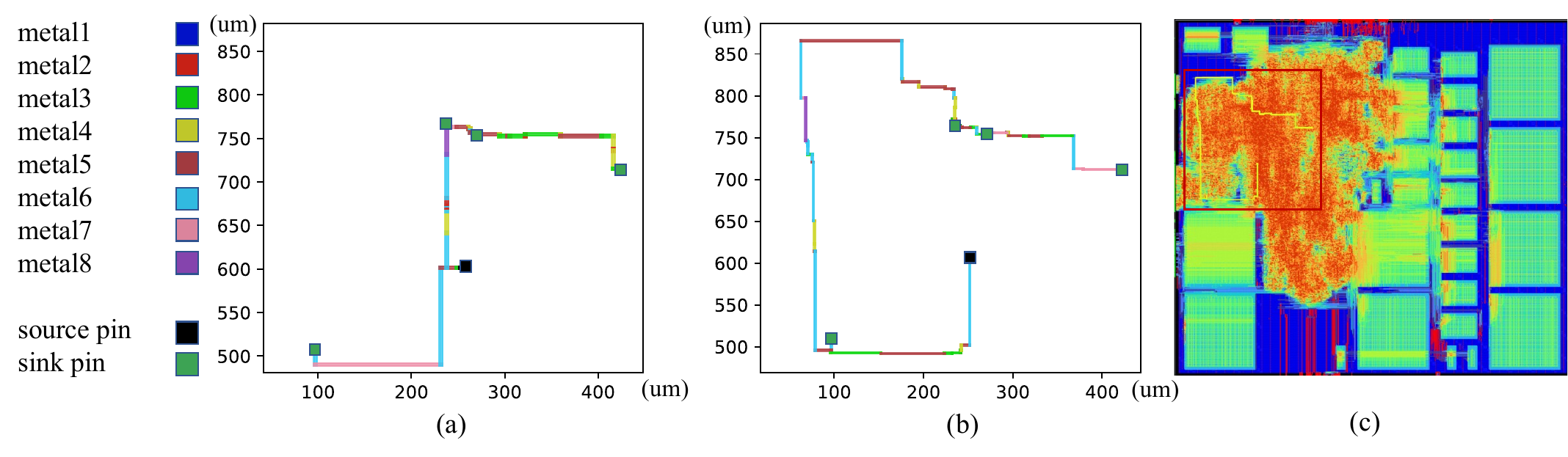}}
\caption{\blueHL{Differences     between the GR guides and post-DR routes of a 5-pin net from swerv\_wrapper 45nm. (a) Route guides in post-GR. (b) Post-DR routes. (c) Net highlighted in congestion map.}}
\label{fig:routes}
\end{figure}

The precise wire routes and wire adjacency relations are available for all nets only after DR is complete. Therefore, post-DR parasitic extraction provides the exact ground-truth parasitics for each route. \blueHL{For a 45nm swerv\_wrapper layout, the GR and DR solutions for a 5-pin net that lies in a high-congestion region are displayed in Figs.~\ref{fig:routes}(a) and (b), respectively. Fig.~\ref{fig:routes}(c) superposes the net over the post-DR congestion map for the design: the red box shows the region displayed in Figs.~\ref{fig:routes}(a) and (b), and the net is highlighted in yellow.  It is clear that the DR solution  of this net chooses a path that is significantly different from that specified in GR, and that this is due to limited available routing resources in the congested region: compared to the GR guides, the DR solution makes a large detour to either bypass the most congested region or \blueHL{overcome pin access issues}. Due to such differences between the post-GR wire route prediction and post-DR wiring topologies of the nets, the timing results based on post-GR parasitics and the ground truth can sometimes be quite different. As a result, GR-stage timing estimates can be inaccurate. Moreover, depending on factors such as changes in the Steiner tree, or the coupling capacitances to neighboring wires, post-GR parasitic estimation may be pessimistic or optimistic, as seen in Fig.~\ref{fig:linear-fit}. This may mislead post-GR timing optimization steps (e.g., buffering/resizing) into performing incorrect or inaccurate optimizations.}
% \footnote{Although the plot in Fig.~\ref{fig:linear-fit} is dominated by optimistic delay predictions, a detailed examination of the data shows a mix of optimistic and pessimistic predictions.}

%As a result, in typical current-day methodologies, although the route guides provide the best parasitic estimates prior to DR, they can be erroneous. This may lead to significant errors in estimating path delays, as visible in the plot on the right in Fig.~\ref{fig:circuit-definitions}.

\subsection{Timing estimation}
\begin{figure}[htbp]
\centerline{\includegraphics[width=0.4\linewidth]{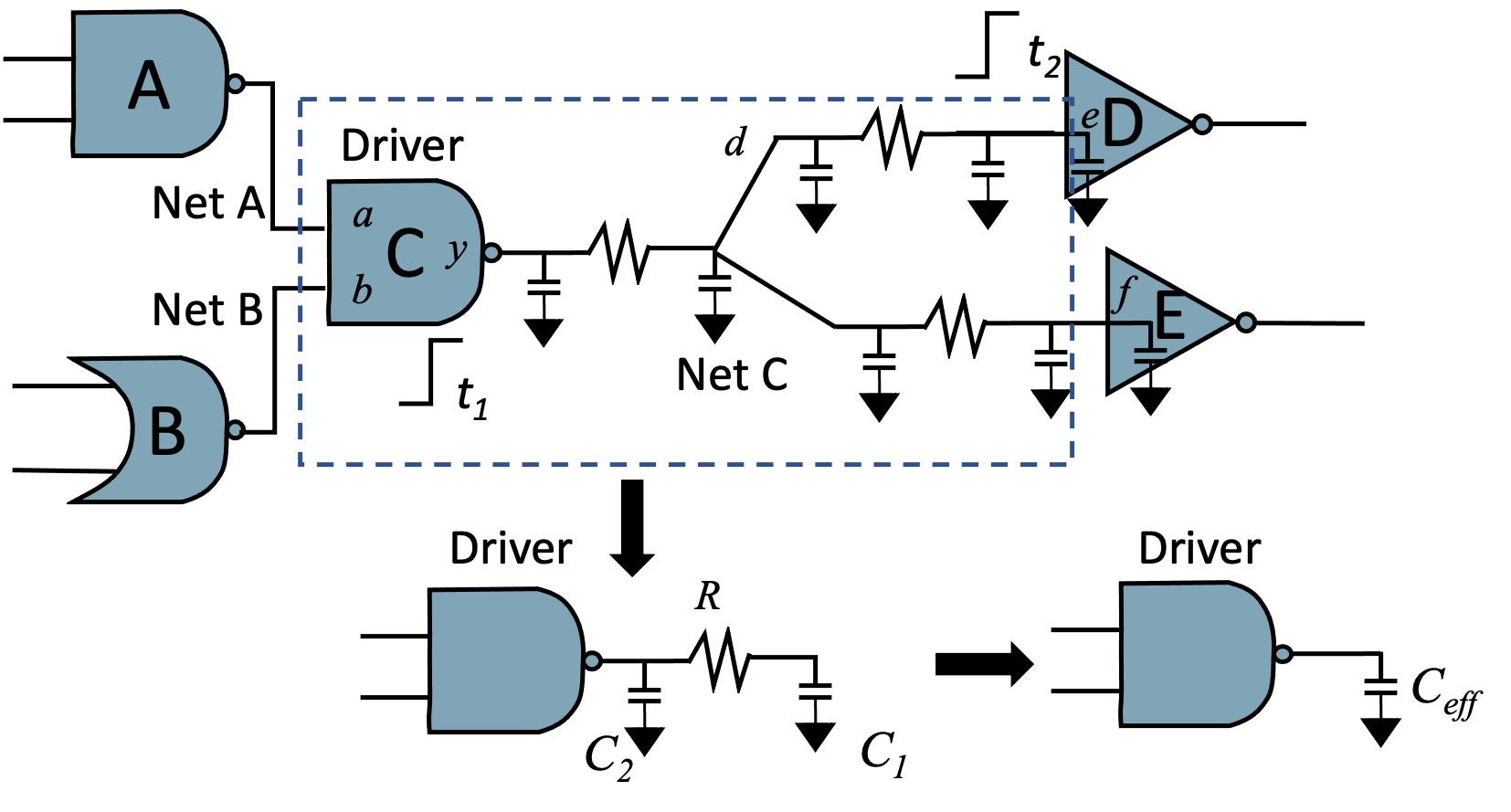}}
%\vspace{-1.0em}
\caption{The RC tree model of a net in a logic stage, and its reduction to an equivalent $\pi$-model.}
%\vspace{-1.5em}
\label{fig:LogicStage}
\end{figure}

A unit operation in static timing analysis is the estimation of the delay of a single logic stage, 
consisting of a gate driving its fanouts through a net, as shown in Fig.~\ref{fig:LogicStage}. The upper part of the figure shows the distributed RC tree for the net driven by gate C. The delay of a logic stage consists of the gate delay and interconnect delay. Given the estimated capacitive load $C_{\mbox{\scriptsize load}}$ at the output of a gate (the ``driving point''), and the transition time $\tau$ at the gate input, the gate delay is expressed through a lookup table (LUT) as
\begin{align}
D_{gate} = f(\tau, C_{\mbox{\scriptsize load}}).
\label{eq:GateDelay}
\end{align}
The gate delay for an intermediate pair of $(\tau,C_{\mbox{\scriptsize load}})$ values that does not map to a LUT entry is computed using interpolation. The transition time at the gate output is estimated in a similar way, using LUTs that have the same axes as gate delay LUTs.

Since a wire is a distributed transmission line, it is modeled by the distributed RC model shown in Fig.~\ref{fig:LogicStage}, which segments the wire and creates lumped approximations for each segment.  If the segments are small, this approximates the derivatives in the differential equation for a transmission line by finite sums. \blueHL{For a short wire, the wire resistance is dominated by the driver, and the wire can be represented by a capacitive load. Thus, the $C_{\mbox{\scriptsize load}}$ model above can be used directly to compute the gate delay, which dominates the stage delay as  the wire delay is negligible in this case.  However, for longer interconnects, resistive shielding effects can significantly impact the delay~\cite{Sapatnekar04}.} For such wires, a segmented RC model cannot directly use the LUT of Eq.~\eqref{eq:GateDelay} as the load is not purely capacitive.  A typical approach to overcome this in timing analysis is by (a)~generating a $\pi$-model reduction for the driving point impedance at the gate output~\cite{O'Brien1989}, with elements $R$, $C_1$, and $C_2$ (Fig.~\ref{fig:LogicStage}, bottom) chosen so that the first three admittance moments of the interconnect match those of the reduced model;  
(b)~using the $\pi$-model to find an effective capacitance, $C_{\mbox{\scriptsize eff}}$, that models resistive shielding; and (3)~using $C_{\mbox{\scriptsize load}}=C_{\mbox{\scriptsize eff}}$ in Eq.~\eqref{eq:GateDelay} to compute the gate delay. Finally, model order reduction techniques are used to compute the transfer function from the driving point to each fanout. Based on the delay and slew at the driving point, the waveform at each fanout is computed, yielding the wire delay and slew.  The sum of the gate and wire delays constitutes the stage delay to the fanout, and the slew at each fanout is used as the input slew for the next logic stage.

\section{DR timing prediction framework}
\label{sec:framework}

\subsection{Timing prediction in routing flow}
\label{sec:MLmodels}

\begin{figure}[tbp]
\centerline{\includegraphics[width=0.6\linewidth]{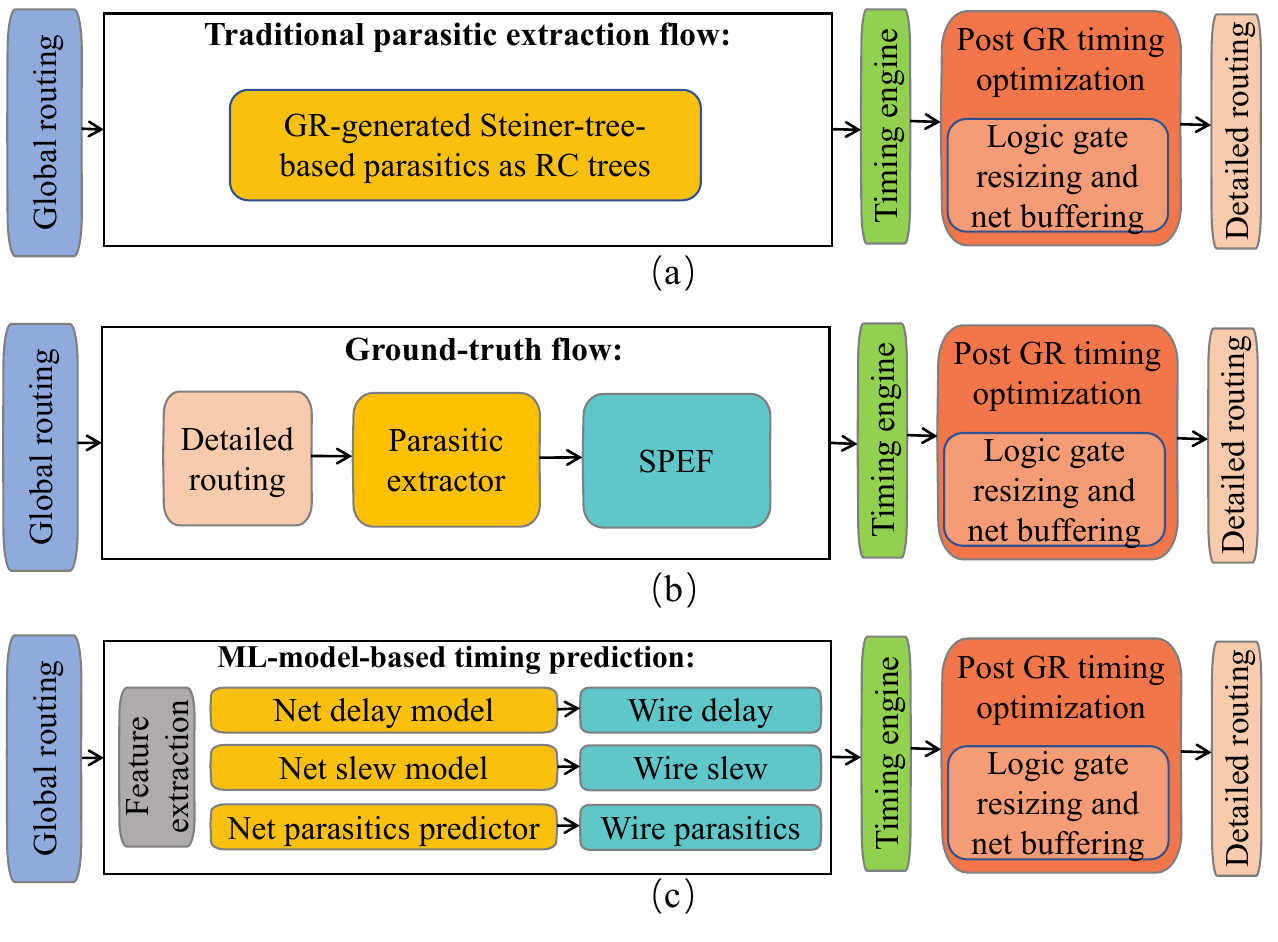}}
\caption{Flows that use different parasitic estimates for post-GR timing optimizations: (a) traditional flow (Steiner tree-based RC estimates), (b) ground-truth (DR followed by parasitic extraction to determine post-DR parasitics), and \blueHL{(c) our flow (fast ML engine for post-DR parasitics and timing prediction).}}
%\vspace{-1.5em}
\label{fig:flows}
\end{figure}

Fig.~\ref{fig:flows}(a) highlights a typical routing flow in physical design, where the GR stage is followed by timing optimization before the final DR stage. These optimizations rely on parasitic estimates from route guides, which use Steiner trees to construct an RC tree network. This results in  timing inaccuracy, relative to the final DR timing (see Fig.~\ref{fig:linear-fit}). In an ideal flow, shown in  Fig.~\ref{fig:flows}(b), performing DR and extracting parasitics would provide accurate timing estimates, but such a flow is impractical due to the high computational expense of DR.

We propose the use of an ML-based flow to predict post-DR timing from features extracted at the post-GR stage in both OpenROAD and a commercial tool flow, as highlighted in Fig.~\ref{fig:flows}(c). Through fast ML inference, the model can rapidly predict post-DR timing estimates without performing time-intensive DR. Our framework leverages three XGBoost-based ML models to predict the following three post-DR metrics:

\noindent
{\bf (i) Source-sink wire delay:} Our ML model is applied on a per-sink basis, and it predicts the delay between the driving point and the sink pins.  For example, in Fig.~\ref{fig:LogicStage}, for net C, the model predicts the wire delay between the driving point (pin $y$ of the driver gate C) and the sink  (pin $e$ of gate D). Similarly, it predicts the source-sink delay between the driver pin $y$ and sink pin $f$.  

\noindent
{\bf (ii) Source-sink wire slew:}
Our ML model is used to predict post-DR wire slew at each sink in the design. We predict source-to-sink wire slews, i.e., the difference in the transition times between the waveforms at the driver pin and at each of the corresponding sink pins. In Fig.~\ref{fig:LogicStage}, the source-sink wire slew is $(t_1-t_2)$, where $t_1$ is the transition time at the driving point pin $y$, and $t_2$ is the transition time at the sink pin $e$.

% \begin{figure}[tbp]
% \centerline{\includegraphics[width=0.6\linewidth]{figs/wirelength_distribution.pdf}}
% \caption{wirelength distribution of swerv\_wrapper 45nm: (a) for design in size of $1.20\times1.09 um^2$ generated in OpenROAD. (b) for design in size of $1.18\times1.18 um^2$ generated in commercial tool.}
% \label{fig:wirelength_distribution}
% \end{figure}

\noindent
\blueHL{{\bf (iii) Wire parasitics:} In the OpenROAD flow, our ML model predicts post-DR $\pi$-model parasitics ($R$, $C_1$, and $C_2$, as shown in Fig.~\ref{fig:LogicStage}). The timing engine uses these three parameters to estimate $C_{\mbox{\scriptsize eff}}$ which is used, in turn, to calculate gate delays. For the commercial flow, the model predicts the post-DR total load capacitance.  In the OpenROAD flow,  the predicted wire delay, wire slew and $\pi$-model parameters are annotated by corresponding commands through Tcl APIs, and the STA engine updates $C_{\mbox{\scriptsize eff}}$ and gate delays based on the $\pi$-model parameters.  In the commercial tool flow, equivalent commands are used to annotate the predicted wire delay and wire slew, and the parasitics are annotated. However, the available APIs do not expose the $\pi$-model parameters, and therefore, we settle for using the total load capacitance, $C_{\mbox{\scriptsize load}}$, for these designs.  For reasons described in Section~\ref{sec:results}, this does not result in significant inaccuracies in the commercial tool flow.}

% \redHL{MOVE THIS AND FIGURE TO RESULTS SECTION: Since the designs generated in commercial tool have fewer long wires as shown in Fig.~\ref{fig:wirelength_distribution}, where we compare the number of nets with length greater than 100$\mu$m of swerv\_wrapper 45nm generated in OpenROAD and commercial tool, using $C_{\mbox{\scriptsize load}}$ does not affect the accuracy of timing estimation too much. Therefore our model predicts $C_{\mbox{\scriptsize load}}$ for gate delay calculation. The annotation of wire delay and wire slew are similar in OpenROAD and in the commercial tool.}

Together, the above ML models (i)-(iii) estimate post-DR circuit delays with the help of an STA engine for annotated delay propagation.  The first two models are directly 
used to annotate wire delays and wire slews in the  timer while the latter is used indirectly in the timer to calculate $C_{\mbox{\scriptsize eff}}$, which is used as $C_{\mbox{\scriptsize load}}$ in \eqref{eq:GateDelay} to compute the gate delay.

\subsection{ML engine}

All three ML models are implemented using XGBoost~\cite{Chen16}, an ensemble learning algorithm based on gradient boosting. XGBoost predicts the target variable using parallel tree boosting, combining estimates from several models including gradient-boosted decision trees. A linear combination of multiple trees is used to describe the complex nonlinear relationship between input and output data. New trees are generated based on previous trees, using gradient descent to minimize a loss function.

\newpage
\subsection{Feature engineering}
\label{sec:features}

\subsubsection{Input features for source-sink delay and slew prediction models}

\begin{figure}[tb]
\center
\includegraphics[width=0.9\linewidth]{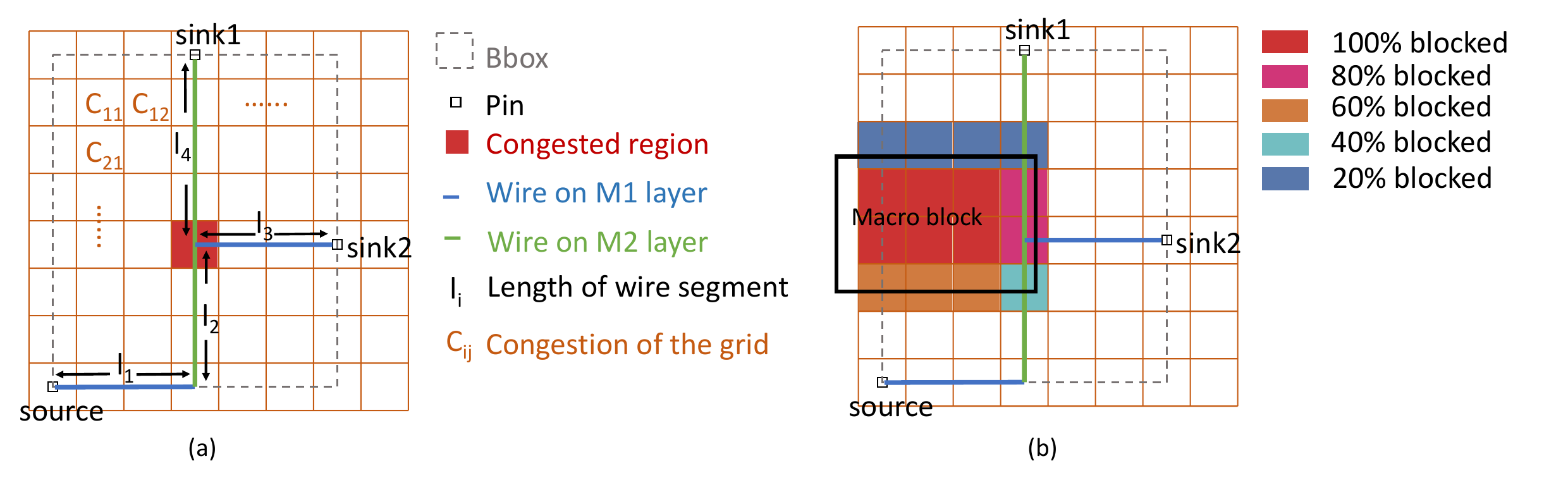}
\caption{A route guide for a three-pin net with a source and two sinks. (a) The net is routed using four wire segments with lengths $l_1$, on M1; $l_2$, on M2; $l_3$, on M1; and $l_4$, on M2. \blueHL{(b) The net traverses two GCells that are 80\% blocked by a macro and two GCells that are 40\% blocked and 20\% blocked.}}
\label{fig:feature-explanation}
\end{figure}

We identify the following set of input features that are used to build the ML model that predicts the source-sink delay and slew of a net. The first set of features described below are derived on a per-net basis, while another set of features, described later, are on a per-sink basis.

\noindent
\textbf{HPWL}: This feature (half-perimeter of the minimum bounding box of the net) is illustrated by the dashed bounding box in Fig.~\ref{fig:feature-explanation}(a). It provides a lower bound on the wirelength. 

\noindent
\textbf{Number of sinks}: It is well known that for nets with numerous sinks, the HPWL can be a significant underestimate. Multi-sink nets show larger discrepancies between post-GR and post-DR wirelengths as they tend to detour and/or have net lengths that exceed the HPWL, as noted above.  We therefore pass this feature, which is extracted from the gate-level netlist, to the ML model. 
%, is typically a distinguishing factor for wire parasitics. 

\noindent
\textbf{Slew at the driving point}: The driving point slew indicates the signal strength at the source pin. This slew is extracted from a timing analysis tool, which uses GR-generated Steiner tree-based parasitics. The slew at the source pin affects both source-sink wire delay and the wire slew: a smaller driving point slew leads to a smaller source-sink delay and slew.

\noindent
\textbf{Congestion estimates}: As illustrated in Fig.~\ref{fig:routes}, congestion is  critical to bridging the discrepancy between GR and DR. Nets whose GR-generated route guides go through congested regions tend to detour in DR, for routability, pin access, and DRC-related reasons. Therefore, congestion estimates are essential for predicting wire detours. As features, we use both the mean and standard deviation of the congestion in all GCells in the bounding box of the net as shown in Fig.~\ref{fig:feature-explanation}(a).

\noindent
\textbf{Rise and fall transitions:} Since the wire delays and slews are different for the rise and fall transitions, we encode the switching direction with a binary-encoded feature (0 for rise and 1 for fall).

The above-listed features are on a per-net basis, i.e., identical for all sinks on the net. Since we predict wire delays and slews on a per-sink basis, we also use the following sink-specific features:

\noindent
\textbf{Source-sink length}: The source-sink length is defined as the total length of the wire segments that connect the driving point to the target sink pin, and is extracted from the GR-generated route guides. 
For the example net in Fig.~\ref{fig:feature-explanation}(a), routed in layers M1 and M2, the source-sink length for sink1 is defined as $(l_1 + l_2+ l_4)$. Since the source-sink length is strongly reflected in to the source-sink delay and slews, this is a critical feature in estimating post-DR wire and slew delays.

\noindent
\textbf{Source-sink R, C}: These features give the total resistance and capacitance, respectively, between the driving point and the target sink. The total resistance [capacitance] is the sum of the products of the segment length and the per-unit resistance $R_{Mi}$ [capacitance $C_{Mi}$] of its assigned layer $Mi$, over all source-to-sink wire segments. In Fig.~\ref{fig:feature-explanation}(a), the total resistance to sink1 is $(l_1\times R_{M1} + l_2\times R_{M2} + l_4\times R_{M2})$, and the total capacitance is $(l_1\times C_{M1} + l_2\times C_{M2} + l_4\times C_{M2})$.

\noindent
\blueHL{\textbf{Routing blockage due to macros}: Since macro blocks squeeze the space available for routing, any nets traversing the GCells blocked by macros must detour around these GCells, with probability that is a function of the degree of blockage; the detouring resutls in increased wire delay, total wire resistance, and total wire capacitance. As shown in Fig.~\ref{fig:feature-explanation}(b), the path from the source to sink1 traverses four GCells that are blocked by macros: two of these GCells are 80\% blocked, one is 40\% blocked, and one is 20\% blocked. Due to limited routing resources, it is likely that the wire connecting source and sink1 will bypass the GCells that are 80\% blocked in the final routing result.  The features associated with the macros must capture the variation in GCell blockage over the regions traversed by a net.  This may be achieved in several ways, e.g., (1)~using the mean of the percentage blockage of the GCells in the bounding box of an entire net; (2)~using the maximum percentage blockage of the GCells in a source-to-sink bounding box; or (3)~using the mean of the percentage blockage of the GCells in the bounding box for a source-sink pair.}

\blueHL{However, for a net with multiple sinks, the first option using the mean blocking percentage of the GCells in the net bounding box can be misleading: it does not estimate the macro impact accurately when one source-sink path goes through a highly blocked GCell (e.g., >80\% blocked) but other source-sink paths traverse less blocked or unblocked GCells. The second option, using the maximum blocking percentage of the GCells in the source-sink bounding box, could mispredict the wire detour when only a small number of GCells are largely blocked by macros. Therefore, we base our approach on the third option, using the mean blocking percentage of all GCells in the source-sink bounding box, to estimate the impact of macro blockage on the wiring inside the source-sink bounding box.}

\blueHL{Furthermore, we observe that even this approach is imperfect.  Notably, a detoured route may traverse GCells outside of the bounding box, and the macro blockages in these GCells outside the bounding box will also affect the routing result. Hence, the average blocking percentage in an expanded bounding box is introduced as a feature to estimate the macro impact, where the expanded bounding box is scaled by scaling factor $\alpha > 1$ over the source-sink bounding box. \ignore{A set of $\alpha$ values are selected, and the mean and max \%error of the nets in a set of 45nm designs that have >80\% average blocking percentage are summarized in Table~\ref{tbl:scaling-factor}. Over this range, we see that $\alpha=2$ shows the largest improvement in the mean percentage error, and is close to the best improvement in the maximum percentage error.}  Empirically, we find that $\alpha=2$ is a good choice and we therefore use an expanded bounding box with $\alpha = 2$, i.e., doubling the length and width of the bounding box, to determine the mean percentage blockage that is used as a feature in the ML model.}

\ignore{
\begin{table}
\centering
\caption{\%error of wire delay at different values of the bounding box scaling factor, $\alpha$.}
\label{tbl:scaling-factor}
\begin{tabular}{|c|c|c|c|c|c|c|c|} 
\hline
Scaling factor $\alpha$ & 1      & 1.25    & 1.5    & 1.75   & 2      & 2.25   & 2.5     \\ 
\hline
mean \%error~        & 1.67\% & 1.54\%~ & 1.49\% & 1.43\% & 1.41\% & 1.51\% & 1.52\%  \\ 
\hline
max \%error          & 7.76\% & 6.81\%  & 6.48\% & 6.35\% & 6.37\% & 6.47\% & 6.52\%  \\
\hline
\end{tabular}
\end{table}
}

\subsubsection{Input features for the load prediction model}

\noindent
For the ML model that predicts the parameters of the $\pi$-model at the driving point for use with the OpenROAD timer, we use the HPWL, number of sinks, congestion estimates and macro blockage estimates as features. We use additional features related to the values of $R$, $C_1$, and $C_2$, generated by applying the O'Brien/Savarino model~\cite{O'Brien1989} to the GR-generated Steiner tree.  \greenHL{Since the timing engine of the commercial tool uses $C_{\mbox{\scriptsize load}}$ instead of $C_{\mbox{\scriptsize eff}}$, and does not provide visibility into the parameters $R$, $C_1$, and $C_2$ of the $\pi$-model, our best option is to use  HPWL, number of sinks, congestion estimates and macro blockage estimates, along with $C_{\mbox{\scriptsize load}}$ based on the GR result, as model features for the commercial tool flow.}

\section{Model training and inference in the physical design flow}
\label{sec:training}

% \subsection{ML engine}

% All three ML models are implemented using XGBoost~\cite{Chen16}, an ensemble learning algorithm based on gradient boosting. XGBoost predicts the target variable using parallel tree boosting, combining estimates from several models including gradient-boosted decision trees. A linear combination of multiple trees is used to describe the complex nonlinear relationship between input and output data. New trees are generated based on previous trees, using gradient descent to minimize a loss function.

\subsection{Ground-truth data generation and model training}

Our ground-truth data is generated using the 
flow highlighted in Fig.~\ref{fig:flows}(b). \blueHL{Since our experiments are performed in different tool flows, we use both a branch of OpenROAD-flow-scripts~\cite{ORFS} and a commercial flow. The ground-truth data is generated for multiple designs implemented in two open-source bulk CMOS technologies (a 45nm technology~\cite{NanGate45}) and a 130nm technology~\cite{SkyWater130}, and one commercial FinFET technology. All of the tested designs are summarized in Table~\ref{tbl:designs}.}

\begin{table}
\centering
\caption{\blueHL{Summary of designs implemented in 45nm, 12nm, and 130nm technology nodes.}}
\label{tbl:designs}
\begin{tabular}{|l|l|l|l|} 
\hline
Design         & Tech                   & \# Nets & Macros  \\ 
\hline
ibex           & \multirow{7}{*}{45nm}  & 17566   & 0       \\ 
\cline{1-1}\cline{3-4}
aes            &                        & 16836   & 0       \\ 
\cline{1-1}\cline{3-4}
jpeg           &                        & 68247   & 0       \\ 
\cline{1-1}\cline{3-4}
dynamic\_node  &                        & 11598   & 0       \\ 
\cline{1-1}\cline{3-4}
swerv\_wrapper &                        & 88490   & 28      \\ 
\cline{1-1}\cline{3-4}
bp\_fe         &                        & 24883   & 11      \\ 
\cline{1-1}\cline{3-4}
bp\_be         &                        & 41973   & 10      \\ 
\hline
swerv\_wrapper & \multirow{2}{*}{12nm}  & 92787   & 28      \\ 
\cline{1-1}\cline{3-4}
coyote         &                        & 272948  & 15      \\ 
\hline
ibex           & \multirow{4}{*}{130nm} & 15307   & 0       \\ 
\cline{1-1}\cline{3-4}
aes            &                        & 15369   & 0       \\ 
\cline{1-1}\cline{3-4}
jpeg           &                        & 59573   & 0       \\ 
\cline{1-1}\cline{3-4}
riscv32i        &                        & 8150    & 0       \\
\hline
\end{tabular}
\end{table}

To increase the diversity of our training data set, we run the ground-truth flow for each design with three different floorplan areas and placement utilization settings. Different utilization settings result in designs with different levels of congestion and consequently in nets that have different lengths due to placements and detours. We extract the features described in Section~\ref{sec:features} and their corresponding ground-truth labels. \blueHL{The training labels for the three ML models are extracted after timing analysis (corresponding to the green box in Fig.~\ref{fig:flows}(b)) including the source-sink wire delays, source-sink wire slews, and load parameters. Each source-sink pair represents a single data point for the source-sink delay and slew at the sink. For the load capacitance predictor, each net in the design is a single data point. \greenHL{The training data is normalized before training; however, in practice we observe that the XGBoost model is not very sensitive to this normalization and performs adequately using skewed data.} The models for designs with macro blocks and without macro blocks are trained separately. For designs without macros, our ground-truth dataset contains 456,661 data points in 45nm technology, 348,429 data points in 12nm technology, and 394,162 data points in 130nm technology. For designs with macros, our ground-truth dataset contains 880,225 data points in 45nm, and 643,793 data points in 12nm.} The pace of ground-truth data generation is slow (1 hour per design, on average) but this is a one-time cost per technology since the trained model can be applied to new designs to rapidly and accurately predict post-DR parasitics and timing.

The model is trained using root mean squared error (RMSE) as the loss function. For the XGBoost regressor~\cite{Chen16}, we use a learning rate 0.01. We choose the maximum tree depth~=~4, the number of estimators~=~900, and the subsampling ratio~=~0.8. The hyperparameter values are chosen based on a full search of a discrete grid on the domain of the hyperparameters: the assignment with the highest score is used for parameter tuning. We find that the models are not sensitive to small changes in the hyperparameter values.

\subsection{ML inference in physical design flow}

The trained ML models are applied to the flow shown in Fig.~\ref{fig:flows}(c).
At the post-GR phase, the ML model features are extracted from the design
data and route guides. Then, the features are fed into the three ML models,
which perform a fast and accurate inference to predict source-to-sink wire
delays and wire slews, \blueHL{as well as $\pi$-model parameters in OpenROAD or effective load capacitance in the commercial tool flow. The predicted estimates are annotated via Tcl APIs in the STA engine: load parameters are used by the timing engine to estimate the gate delay, while the source-sink wire delays and slews are directly used as the net delays and net transition times.} The timer performs an update to propagate 
these annotated wire delays and gate delay estimates. The new ML-predicted 
timing estimates are used by the timing optimizer to perform gate resizing
and buffer insertions which fix setup, hold,  maximum slew, maximum fanout, and maximum load violations.

As confirmed in our experimental studies (Seciton~\ref{sec:results} below), the ML-based inference flow provides an accurate estimate of post-DR timing without performing time-intensive DR. These estimates are useful for efficient buffering and resizing, as the optimizer now has improved knowledge of the truly (post-DR) critical paths.

\subsection{Feature sensitivity}

\begin{figure}[t]
\begin{subfigure}{0.7\textwidth}
\includegraphics[width=\linewidth]{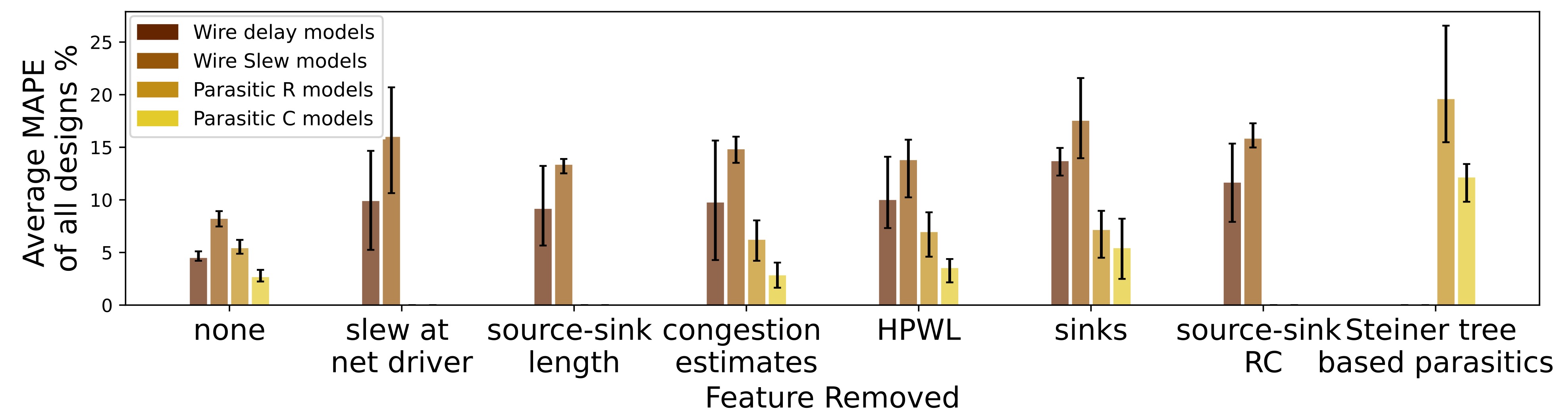}
\caption{Average MAPE on designs without macros in 45nm.}\label{fig:feature-importance-nomacro-45nm}
\end{subfigure}
\begin{subfigure}{0.9\textwidth}
\includegraphics[width=\linewidth]{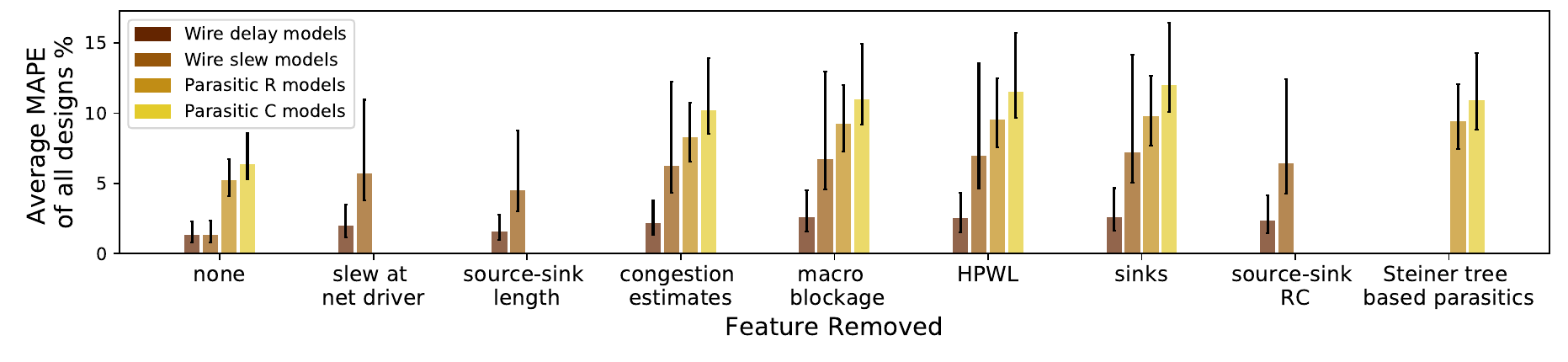}
\caption{Average MAPE on designs with macros in 45nm.}\label{fig:feature-importance-45nm}
\end{subfigure}
\begin{subfigure}{0.9\textwidth}
\includegraphics[width=\linewidth]{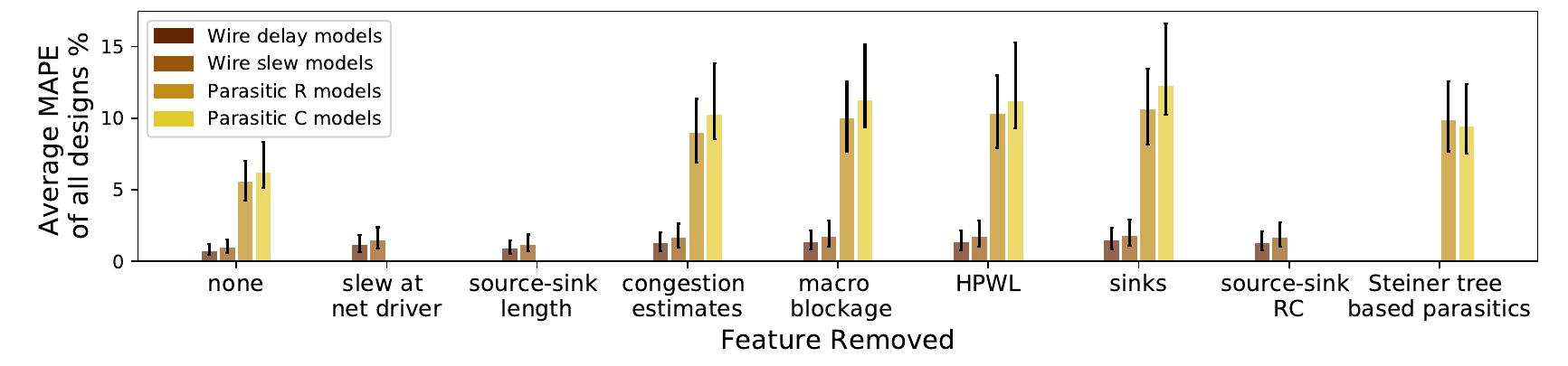}
\caption{Average MAPE on designs with macros in 12nm.}\label{fig:feature-importance-12nm}
\end{subfigure}
\caption{\blueHL{Feature sensitivity analysis showing the average mean absolute percentage error (MAPE) (y-axis), for each removed feature (x-axis), for all 45nm designs and all 12nm designs.}}
\label{fig:feature-importance}
\end{figure}

\begin{figure}[tbp]
\begin{subfigure}{0.4\textwidth}
\includegraphics[width=\linewidth]{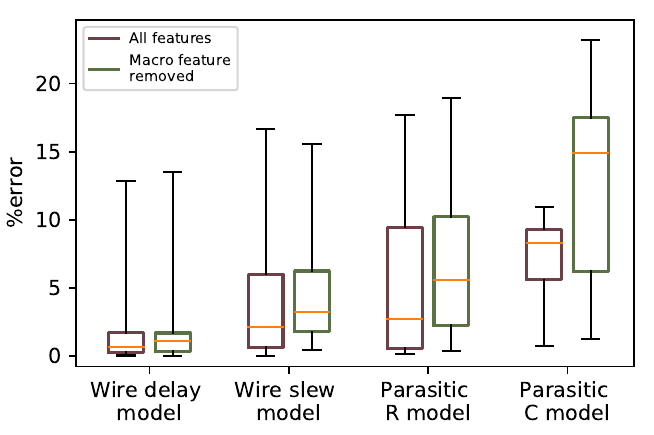}
\caption{Macro feature sensitivity for 45nm designs.}\label{fig:macro-feature-importance-45nm}
\end{subfigure}
\begin{subfigure}{0.4\textwidth}
\includegraphics[width=\linewidth]{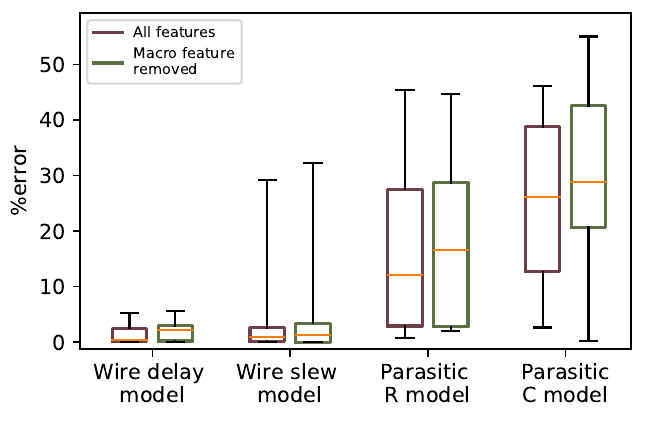}
\caption{Macro feature sensitivity for 12nm designs.}\label{fig:macro-feature-importance-12nm}
\end{subfigure}

\caption{\blueHL{Macro feature sensitivity for wires that have area in a bounding box that with >80\% blockage.}}
\label{fig:macro-feature-importance}
\end{figure}

To demonstrate that each selected feature is indispensable to the model, we perform a sensitivity analysis for each
feature, for each of the four models (wire delay, wire slew, parasitic R, and parasitic C) from Section~\ref{sec:MLmodels}.  Note that not all models use all features; therefore, there are missing bars for certain features in the figure. \greenHL{Fig.~\ref{fig:feature-importance} performs an ablation study on the set of features, removing one feature at a time and examining its impact on the accuracy of the ML model for designs with macros. The analysis of the ML model for designs without macros was conducted in our previous conference version~\cite{Chhabria22}.} For example, for source-sink wire delay prediction, we measure the test accuracy for wire delay prediction models, each trained by removing one specific feature at a time. Similar experiments are also performed on the wire slew model, parasitic R model, and parasitic C model. \greenHL{The y-axes of Figs.~\ref{fig:feature-importance}(a)-(c) respectively show the average of the mean average percentage errors (MAPEs) over all 45nm designs without macros, all 45nm designs with macros, and all 12nm designs with macros. The x-axis of each figure lists the feature that has been removed.}  The figure is annotated with an error bar that shows the maximum and minimum \%error across all the designs.  We find that the model has the best test accuracy when all the features are selected. Thus, each feature contributes to improving the accuracy of the model.  \blueHL{For the 12nm model, which has lower parasitics, the errors approximately double with removal of slew and source-sink features, but since the baseline (``None'') is low, the error after removal is also low.  Nevertheless, we keep these features in the model to enable generality.}
% \redfn{The last sentence isn't true for all features in 12nm -- can we explain why?\blueHL{Parasitic bars dominate in 12nm figure. When we zoom in the figure to see bars for wire delay and wire slew, it's also true.}}

\blueHL{We further examine the macro feature sensitivity in Fig.~\ref{fig:macro-feature-importance} for the four ML models. The x-axis of the figure lists the models trained with the macro features, and the y-axis highlights the average percentage error of the nets in all 45nm designs and all 12nm designs.
Nets whose expanded bounding box has a large overlap with macro blockages are much more greatly affected by macro features than those that are more distant. For these nets, 
% Since macro features have a larger impact on the nets that have large area blocked by macros in their bounding box than the nets more distant from macros, 
we plot data for source-sink pairs that have 80\% of their source-sink bounding box area blocked by macros. Fig.~\ref{fig:macro-feature-importance} shows that introducing macro features improves the average absolute \%error from (1.87\%, 3.24\%, 5.58\%, 14.90\%) to (1.41\%, 2.11\%, 2.72\%, 8.31\%) for 45nm designs, where the four percentages correspond to average absolute \%error of the four models in the figure. For 12nm designs, the corresponding numbers in the average absolute \%error show an improvement from (2.17\%, 1.21\%, 16.56\%, 28.90\%) to (0.45\%, 0.94\%, 12.07\%, 26.17\%).}
\section{Experimental setup and evaluation}
\label{sec:results}

Our experiments are performed on benchmarks from OpenROAD across three technologies: two open technologies, NanGate 45nm~\cite{NanGate45} and SkyWater 130nm~\cite{SkyWater130}, and one 12nm commercial technology.  We perform physical design using two flows: with OpenROAD and a commercial EDA tool flow. Our target designs fall into two classes: those without macros (IBEX, AES, JPEG and Dynamic Node in 45nm; IBEX, AES, JPEG and RISCV32I in 130nm; and IBEX, AES and JPEG in 12nm) and those with macros (swerv\_wrapper, bp\_fe and bp\_be in 45nm; and swerv\_wrapper and coyote in 12nm).  \blueHL{As compared to the preliminary version of this paper~\cite{Chhabria22}, where the results were shown only on bulk nodes, we now show results here on a FinFET technology node; we apply the approach using not only OpenROAD, but also using a commercial tool flow; and we expand the algorithm to address designs with macros. In addition, we test the ability of our models to generalize with respect to clock periods by running inference on unseen designs generated by using different clock constraints, and we also test the robustness of our models to noise by adding Gaussian noise to training datasets. As detailed below, these comprehensive experiments confirm the general applicability of our approach.}

\greenHL{The ML models for designs with and without macros are trained separately. In Table~\ref{tbl:models-with-without-macros}, we compare the prediction results of {\em one shared} model for both designs with and without macros, versus the results of {\em two separate} models for designs with macros and without macros. For the training of the shared model, we leave out each test design from all designs in 45nm and use the remaining designs for training. The table shows that the model trained specifically for designs with macros or designs without macros has better accuracy than a shared model trained for all designs. We explain this by observing that macro-induced detours and blockages 
pose specific challenges due to the 100\% blockage and the large sizes of the macros, which cannot be easily captured in the shared model. Therefore, building separate models is better than using a shared model.} For designs without macros, the models tested on each design are trained using data from the remaining designs without macros in the same technology. \blueHL{For designs with macros, we are handicapped by the limited number of available designs: i.e., only three designs in the 45nm node and two in the 12nm node. Therefore, the models are trained by using the data from other designs with macros at various utilization settings, and also from the same design, but for different utilization settings. Thus, even with this limited dataset, we ensure that the ML model is evaluated for an unseen design having a different utilization than the set of designs in the training set.}

\begin{table}
\centering
\caption{\greenHL{Prediction error of the shared ML model, and the separate ML models for designs with macros and designs without macros in 45nm.}}
\label{tbl:models-with-without-macros}
\resizebox{\linewidth}{!}{
\begin{tabular}{|c|c|c|c|c|c|c|c|c|c|c|c|c|c|c|} 
\hline
\multirow{3}{*}{\textbf{Designs}} & \multirow{3}{*}{\begin{tabular}[c]{@{}c@{}}\textbf{Clock}\\\textbf{period}\end{tabular}} & \multirow{3}{*}{\textbf{\# Macros}} & \multicolumn{6}{c|}{\textbf{ Wire delay}}                                                           & \multicolumn{6}{c|}{\textbf{Wire slew}}                                                                                \\ 
\cline{4-15}
                                  &                                                                                          &                                     & \multicolumn{3}{c|}{\textbf{ML model (separate)}} & \multicolumn{3}{c|}{\textbf{ML Model (shared)}} & \multicolumn{3}{c|}{\textbf{\textbf{ML model (separate)}}} & \multicolumn{3}{c|}{\textbf{\textbf{ML Model (shared)}}}  \\ 
\cline{4-15}
                                  &                                                                                          &                                     & Mean   & Max      & Stdev.                        & Mean   & Max      & Stdev.                      & Mean   & Max     & Stdev.                                  & Mean    & Max     & Stdev.                                \\ 
\hline
swerv\_wrapper                    & 2.5ns                                                                                    & 28                                  & 0.84\% & 115.19\% & 2.78\%                        & 2.95\% & 149.41\% & 4.83\%                      & 0.04\% & 19.81\% & 0.34\%                                  & 0.55\%  & 22.72\% & 0.61\%                                \\ 
\hline
bp\_fe                            & 2.2ns                                                                                    & 11                                  & 0.64\% & 19.14\%  & 0.91\%                        & 2.76\% & 32.34\%  & 2.78\%                      & 0.04\% & 4.42\%  & 0.13\%                                  & 0.47\%  & 6.46\%  & 0.65\%                                \\ 
\hline
bp\_be                            & 2.8ns                                                                                    & 10                                  & 2.16\% & 34.58\%  & 3.42\%                        & 2.93\% & 48.90\%  & 3.06\%                      & 0.07\% & 9.15\%  & 0.25\%                                  & 0.49\%  & 8.85\%  & 0.59\%                                \\ 
\hline
dynamic\_node                     & 1.0ns                                                                                    & 0                                   & 4.21\% & 39.82\%  & 5.00\%                        & 6.60\% & 53.29\%  & 7.58\%                      & 8.19\% & 39.96\% & 8.18\%                                  & 8.96\%  & 67.78\% & 7.60\%                                \\ 
\hline
ibex                              & 2.0ns                                                                                    & 0                                   & 4.30\% & 38.38\%  & 5.99\%                        & 4.93\% & 35.33\%  & 3.54\%                      & 8.94\% & 38.32\% & 7.35\%                                  & 9.51\%  & 50.48\% & 7.06\%                                \\ 
\hline
aes                               & 0.8ns                                                                                    & 0                                   & 5.12\% & 39.90\%  & 7.46\%                        & 7.35\% & 52.59\%  & 10.43\%                     & 8.23\% & 39.98\% & 8.17\%                                  & 14.88\% & 65.03\% & 12.62\%                               \\ 
\hline
jpeg                              & 1.4ns                                                                                    & 0                                   & 4.13\% & 39.68\%  & 5.56\%                        & 8.41\% & 54.71\%  & 8.01\%                      & 7.47\% & 39.97\% & 9.09\%                                  & 10.54\% & 46.30\% & 8.12\%                                \\
\hline
\end{tabular}
}
\end{table}

% removing this as per Vidya-Wenjing conversation. Cause we should not be sampling 30\%. We should be using all data.
%We divide the wirelength of the training designs into bins and sample 30\% of the nets in each bin.}\redfn{The last sentence is unclear to me. I understand this is some kind of sampling for diversity in the training dataset. But it's unclear why you would divide a training dataset and not just use the whole thing. Unless this is the entire generated dataset and not training alone.}

Training and inference for the ML model are implemented in Python 3.6 and performed on a machine with Intel Xeon Silver 4214 CPU @2.2GHz and
NVIDIA A100 PCIe 40GB GPU. For the OpenROAD flow, the predicted parasitics, wire delays, and wire slews are annotated into the timing engine by modifying the OpenROAD~\cite{OpenROAD} and OpenSTA~\cite{OpenSTA} source code.  For the commercial flow, the predicted values are annotated into the timing engine through available Tcl commands in the commercial tool flow.

We evaluate our ML-based flow against the traditional (``Trad.'') and ground-truth-based flows for (i)~accuracy, and (ii)~impact on post-DR outcomes -- worst slack (WS), total negative slack (TNS), runtime and congestion. To highlight the importance of incorporating macro-based features, we also compare the results from the new flow, using a macro-based ML model, to a flow similar to our preliminary work~\cite{Chhabria22}, where the models are trained without macro features. 
% ~\redHL{ (2) We need results that compare to MLCAD. Highlight which designs which have macros and how much of a difference macros make. (3) Your text states that we are also going to be comparing buffering and resizing and runtime; currently no mention of any of this in the tables figures or text. See the comment in my email. Runtimes is good. Do not think the others make sense to me.}

\subsection{Model accuracy evaluation}

We analyze the accuracy of the ML models at both the net/sink level (the ML model performs inference on a per-net/per-sink basis) and at the path level. It is critical to evaluate the predicted timing at both levels, as net-level errors have the potential to accumulate or cancel during delay propagation. 

\subsubsection{Net-level accuracy}

\begin{table}
\centering
\caption{\blueHL{Evaluation of the ML models using mean, maximum, and standard deviation of the absolute \%error as metrics.}}
\label{tbl:ml-sink-accuracy2}
\begin{adjustbox}{angle=90}
\resizebox{\textheight}{!}{
\begin{tabular}{|c|l|c|c|c|c|c|c|c|c|c|c|c|c|c|c|c|c|c|c|c|c|} 
\hline
\multirow{3}{*}{\textbf{Tool}} & \multicolumn{1}{c|}{\multirow{3}{*}{\textbf{Designs}}} & \multirow{3}{*}{\begin{tabular}[c]{@{}c@{}}\textbf{Clock}\\\textbf{period}\end{tabular}} & \multirow{3}{*}{\begin{tabular}[c]{@{}c@{}}\textbf{Technology}\\\textbf{node}\end{tabular}} & \multicolumn{6}{c|}{\textbf{ Wire delay}}                                            & \multicolumn{6}{c|}{\textbf{Wire slew}}                                         & \multicolumn{6}{c|}{\textbf{ Path delay}}                                            \\ 
\cline{5-22}
                               &                                   &                                                                                          &                                                                                             & \multicolumn{3}{c|}{\textbf{ML-based}}      & \multicolumn{3}{c|}{\textbf{GR-based}} & \multicolumn{3}{c|}{\textbf{ML-based}} & \multicolumn{3}{c|}{\textbf{GR-based}} & \multicolumn{3}{c|}{\textbf{ML-based}}     & \multicolumn{3}{c|}{\textbf{GR-based}}  \\ 
\cline{5-22}
                               &                                   &                                                                                          &                                                                                             & Mean    & Max                      & Stdev. & Mean    & Max      & Stdev.            & Mean   & Max     & Stdev.              & Mean    & Max     & Stdev.             & Mean   & Max     & Stdev.                  & Mean   & Max     & Stdev.               \\ 
\hline
\multirow{13}{*}{OpenROAD}     & swerv\_wrapper*                   & 2.5ns                                                                                    & \multirow{7}{*}{45nm}                                                                       & 0.84\%  & 115.19\%                 & 2.78\% & 4.27\%  & 149.41\% & 7.28\%            & 0.04\% & 19.81\% & 0.34\%              & 0.20\%  & 40.63\% & 1.22\%             & 0.34\% & 4.62\%  & 0.30\%                  & 1.48\% & 7.54\%  & 0.69\%               \\ 
\cline{2-3}\cline{5-22}
                               & bp\_fe*                           & 2.2ns                                                                                    &                                                                                             & 0.64\%  & \textcolor{red}{19.14\%} & 0.91\% & 3.63\%  & 15.68\%  & 4.65\%            & 0.04\% & 4.42\%  & 0.13\%              & 0.24\%  & 5.83\%  & 0.98\%             & 0.65\% & 1.91\%  & 0.41\%                  & 0.69\% & 2.47\%  & 0.52\%               \\ 
\cline{2-3}\cline{5-22}
                               & bp\_be*                           & 2.8ns                                                                                    &                                                                                             & 2.16\%  & 34.58\%                  & 3.42\% & 3.00\%  & 55.64\%  & 3.88\%            & 0.07\% & 9.15\%  & 0.25\%              & 0.12\%  & 10.45\% & 0.38\%             & 0.63\% & 2.21\%  & \textcolor{red}{0.45\%} & 0.80\% & 2.57\%  & 0.34\%               \\ 
\cline{2-3}\cline{5-22}
                               & dynamic\_node                     & 1.0ns                                                                                    &                                                                                             & 4.21\%  & 39.82\%                  & 5.00\% & 8.43\%  & 51.33\%  & 6.45\%            & 8.19\% & 39.96\% & 8.18\%              & 10.56\% & 51.51\% & 10.54\%            & 0.57\% & 5.84\%  & 0.32\%                  & 2.35\% & 10.76\% & 2.43\%               \\ 
\cline{2-3}\cline{5-22}
                               & ibex                              & 2.0ns                                                                                    &                                                                                             & 4.30\%  & 38.38\%                  & 5.99\% & 6.54\%  & 49.47\%  & 9.62\%            & 8.94\% & 38.32\% & 7.35\%              & 11.52\% & 49.40\% & 9.47\%             & 0.70\% & 6.64\%  & 0.42\%                  & 2.11\% & 10.66\% & 1.18\%               \\ 
\cline{2-3}\cline{5-22}
                               & aes                               & 0.8ns                                                                                    &                                                                                             & 5.12\%  & 39.90\%                  & 7.46\% & 11.60\% & 68.43\%  & 9.62\%            & 8.23\% & 39.98\% & 8.17\%              & 10.61\% & 54.53\% & 10.53\%            & 0.82\% & 4.21\%  & 0.30\%                  & 2.63\% & 11.04\% & 1.72\%               \\ 
\cline{2-3}\cline{5-22}
                               & jpeg                              & 1.4ns                                                                                    &                                                                                             & 4.13\%  & 39.68\%                  & 5.56\% & 5.32\%  & 57.15\%  & 7.17\%            & 7.47\% & 39.97\% & 9.09\%              & 9.63\%  & 51.52\% & 11.72\%            & 2.82\% & 10.60\% & 2.03\%                  & 6.07\% & 29.78\% & 4.28\%               \\ 
\cline{2-22}
                               & swerv\_wrapper*                    & 1.2ns                                                                                    & \multirow{2}{*}{12nm}                                                                       & 0.87\%  & 22.80\%                  & 0.93\% & 4.26\%  & 44.95\%  & 5.76\%            & 0.04\% & 29.81\% & 0.17\%              & 0.70\%  & 71.75\% & 3.85\%             & 1.15\% & 7.61\%  & 1.26\%                  & 5.75\% & 14.91\% & 4.77\%               \\ 
\cline{2-3}\cline{5-22}
                               & coyote*                            & 3.2ns                                                                                    &                                                                                             & 0.60\%  & 26.58\%                  & 0.88\% & 3.47\%  & 39.65\%  & 4.73\%            & 0.22\% & 40.13\% & 0.62\%              & 1.97\%  & 51.34\% & 4.51\%             & 1.09\% & 2.65\%  & 0.27\%                  & 1.23\% & 7.73\%  & 1.51\%               \\ 
\cline{2-22}
                               & ibex                              & 16.0ns                                                                                   & \multirow{4}{*}{130nm}                                                                      & 3.35\%  & 21.21\%                  & 6.33\% & 4.32\%  & 27.34\%  & 8.16\%            & 4.15\% & 21.54\% & 5.45\%              & 5.35\%  & 36.77\% & 7.03\%             & 0.64\% & 8.17\%  & 1.05\%                  & 1.09\% & 4.99\%  & 1.13\%               \\ 
\cline{2-3}\cline{5-22}
                               & aes                               & 5.4ns                                                                                    &                                                                                             & 11.15\% & 39.05\%                  & 8.03\% & 25.37\% & 72.34\%  & 10.35\%           & 2.46\% & 29.18\% & 3.15\%              & 3.17\%  & 65.61\% & 4.06\%             & 0.47\% & 3.22\%  & 0.40\%                  & 1.48\% & 7.77\%  & 0.53\%               \\ 
\cline{2-3}\cline{5-22}
                               & jpeg                              & 7.8ns                                                                                    &                                                                                             & 4.17\%  & 37.83\%                  & 6.51\% & 14.38\% & 48.76\%  & 8.39\%            & 5.84\% & 39.97\% & 6.22\%              & 7.53\%  & 67.52\% & 8.02\%             & 0.82\% & 6.87\%  & 0.65\%                  & 3.20\% & 12.65\% & 2.31\%               \\ 
\cline{2-3}\cline{5-22}
                               & riscv32i                          & 9.6ns                                                                                    &                                                                                             & 1.54\%  & 3.44\%                   & 0.67\% & 2.99\%  & 8.43\%   & 0.86\%            & 1.15\% & 2.58\%  & 0.83\%              & 1.48\%  & 3.33\%  & 0.84\%             & 1.08\% & 2.85\%  & 0.42\%                  & 4.13\% & 19.90\% & 3.27\%               \\ 
\hline
\multirow{5}{*}{\begin{tabular}[c]{@{}c@{}}{Commercial}\\tool\end{tabular}}        
                               & swerv\_wrapper*                   & 2.2ns                                                                                    & \multirow{3}{*}{45nm}                                                                       & 7.21\%  & 63.59\%                  & 6.50\% & 29.80\% & 70.56\%  & 11.56\%           & 0.24\% & 9.14\%  & 0.71\%              & 1.46\%  & 50.93\% & 5.28\%             & 0.35\% & 5.49\%  & 0.53\%                  & 1.00\% & 5.94\%  & 0.60\%               \\ 
\cline{2-3}\cline{5-22}
                               & bp\_fe*                           & 1.6ns                                                                                    &                                                                                             & 5.88\%  & 43.79\%                  & 4.96\% & 24.41\% & 51.07\%  & 9.26\%            & 0.36\% & 6.08\%  & 0.79\%              & 3.73\%  & 37.84\% & 8.83\%             & 0.54\% & 4.81\%  & 0.46\%                  & 1.64\% & 7.05\%  & 0.67\%               \\ 
\cline{2-3}\cline{5-22}
                               & bp\_be*                           & 2.0ns                                                                                    &                                                                                             & 6.09\%  & 57.99\%                  & 5.46\% & 25.84\% & 56.00\%  & 9.29\%            & 0.36\% & 19.30\% & 0.83\%              & 3.04\%  & 99.94\% & 7.84\%             & 0.60\% & 5.08\%  & 0.47\%                  & 1.65\% & 6.28\%  & 0.50\%               \\ 
\cline{2-22}
                               & swerv\_wrapper*                   & 1.0ns                                                                                    & \multirow{2}{*}{12nm}                                                                       & 6.24\%  & 36.47\%                  & 1.27\% & 12.30\% & 56.30\%  & 2.78\%            & 0.36\% & 26.26\% & 0.95\%              & 0.85\%  & 79.14\% & 2.78\%             & 3.52\% & 8.28\%  & 2.29\%                  & 5.98\% & 13.69\% & 3.24\%               \\ 
\cline{2-3}\cline{5-22}
                               & coyote*                           & 3.2ns                                                                                    &                                                                                             & 5.22\%  & 83.62\%                  & 1.05\% & 15.42\% & 98.70\%  & 2.99\%            & 3.68\% & 59.35\% & 0.88\%              & 11.75\% & 93.02\% & 2.86\%             & 2.22\% & 7.36\%  & 1.74\%                  & 4.67\% & 10.85\% & 2.87\%               \\
\hline
\end{tabular}
}
\end{adjustbox}
\end{table}

% \begin{figure}[tbp]
% \begin{subfigure}{0.49\textwidth}
% \includegraphics[width=\linewidth]{figs/trumpet_45nm.jpg}
% \caption{Nets of all designs in 45nm}\label{fig:ml-accuracy-net-level-scatter-a}
% \end{subfigure}
% \begin{subfigure}{0.49\textwidth}
% \includegraphics[width=\linewidth]{figs/trumpet_12nm.jpg}
% \caption{Nets of all designs in 12nm}\label{fig:ml-accuracy-net-level-scatter-b}
% \end{subfigure}
% \caption{Error distribution over ground truth for wire delay and wire slew predition in 45nm and 12nm. Ground truth values are the reference for \% error. ~\redHL{ This related to my masking errors question. Also, how low of a golden value should we not care about? For e.g., is 0.05ns considered low? Need some sense of a reference that must be explained in text. For e.g., 50ps how much percentage of the clock period? Or is how much percentage of the stage delay because before 50ps the errors are large? }}
% \label{fig:ml-accuracy-net-level-scatter}
% \end{figure}

\begin{figure}[tbp]
\centerline{\includegraphics[width=0.99\linewidth]{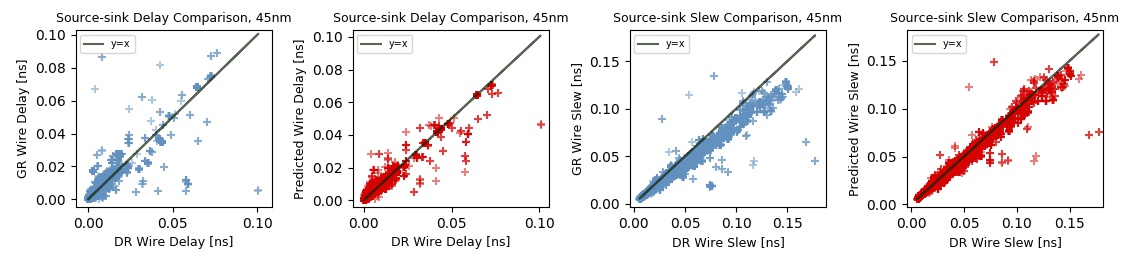}}
\caption{Wire delay and slew comparisons for swerv\_wrapper 45nm.}
%\blueHL{Need to confirm if the second row is what we want to show the improvement of ML prediction}\redHL{(1)~Can remove the second row of figures [this addresses Vidya's comment here (``Need to confirm...''). See my comments in a footnote in the writeup.}
\label{fig:delay_slew_comparison}
\end{figure}

Table~\ref{tbl:ml-sink-accuracy2} summarizes the metrics for the ML models and the accuracy for 45nm designs and 12nm designs in both OpenROAD and the commercial tool flow, and 130nm designs in OpenROAD, evaluated on designs with and without macros. We use the mean, maximum, and standard deviation of the absolute \%error as metrics for evaluation, where the absolute percentage error is the absolute difference between the ground-truth and the predicted value with respect to a reference. The references that we use are the stage delay for evaluating wire delay; the ground-truth sink slew for wire slew; and the ground-truth labels for parasitics.

\blueHL{The table shows better mean \%error and maximum \%error metrics for our ML-based wire delay and wire slew predictions compared to GR Steiner-tree-based estimation from both OpenROAD and the commercial tool flow for most designs. In general, the ML-based model shows significant reductions in the \%error. 
% While the maximum \%error of ML-based wire delay of bp\_fe, 45nm in OpenROAD is larger than that for GR-based wire delay, and the maximum\%error of swerv\_wrapper 45nm in OpenROAD is greater than \%100, we verify that these errors are from shorts nets that have negligible wire delay and small stage delay 
%(e.g., for swerv\_wrapper, the case with 115.19\% error is from a wire with 0.009ns wire delay and 0.018ns stage delay.
In only two cases, denoted in red, the error for our approach is worse than the GR-based approach, but we have verified that these errors are from short nets that have negligible wire delay and small stage delay, and therefore do not affect the critical path in the circuit.}

\blueHL{In some cases in the table, the maximum \%error is very large (e.g., over 100\% for swerv\_wrapper 45nm).  We have confirmed that this error can also be attributed to a short wire that also has a small stage delay as a reference. 
% Fig.~\ref{fig:ml-accuracy-net-level-scatter} shows the signed \%error distribution overs ground-truth wire delay and wire slew values.
Scatter plots of the wire delay and wire slew comparisons in ns, without normalization, are presented in Fig.~\ref{fig:delay_slew_comparison}, showing a more complete picture of the match between ML prediction and the ground truth, relative to the GR-based approach.  It can be seen that the match for the ML predictor is considerably better than for the GR-based approach (even the outlier at the right of the second plot is significantly closer to the $x=y$ line than for the GR-based case).
% The maximum\%error over 100\% from swerv\_wrapper 45nm is from small wire that also has small stage delay as reference. Fig.~\ref{fig:ml-accuracy-net-level-scatter} shows the signed \%error distribution overs ground-truth wire delay and wire slew values. The reference for \%error of wire delay is ground-truth value instead of the stage delay used in Table~\ref{tbl:ml-sink-accuracy}, and the reference for \%error of wire slew is the ground-truth of source-sink slew change predicted directly by wire slew model instead of DR sink slew used in Table~\ref{tbl:ml-sink-accuracy}. 
Since the large errors are only in cases where the ground-truth wire delay and wire slew are very small, the path delay is not greatly affected by these errors. 
From the table and the scatter plots, it can be seen that the standard deviation of \%error from ML models is also lower, indicating that few nets have larger errors. For our application of post-GR timing optimization, these accuracy levels are sufficient to realize the benefit of the ML models.}
% \redfn{This was too much explanation and too many plots to explain a relatively simple issue. I have simplified the presentation by removing the trumpet plots. I think it's enough to show the table, then say that the large errors are due to short stage delays that don't affect the path delay by much, and demonstrate it using the $x=y$ plots.~\blueHL{Agree}}\redfn{Please check this wording and remove redHL if ok.  A reviewer could argue that this is still a problem because it does not identify short paths very well and could lead to hold time constraints, but that is really not a big issue: hold time constraints on short paths are better solved by adding more buffers along short paths rather than increasing wire delays.  I removed the text to avoid drawing too much attention to this issue. It should also be okay to remove the second row (last two figures) of Fig.~\ref{fig:delay_slew_comparison}.~\blueHL{Done}}
 % ~\redHL{(2) Big concern: Even with using stage delay as a reference, the maximum\%erros are over 100\%?  In some cases it's even 150\%.  Please confirm what the reference is? Is it ground-truth wire delay or stage delay? }

\begin{figure}[tbp]
\centerline{\includegraphics[width=0.6\linewidth]{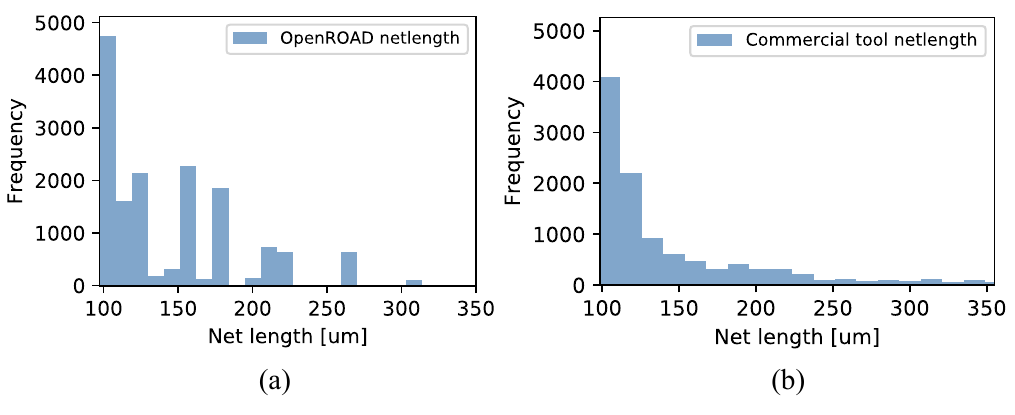}}
\caption{\blueHL{Wirelength distribution of swerv\_wrapper 45nm: (a)~for a design with size of $1.20\times1.09 mm^2$ generated in OpenROAD, and (b)~for a design with size of $1.18\times1.18 mm^2$ generated in a commercial tool flow.}}
\label{fig:wirelength_distribution}
\end{figure}

\blueHL{It is worth noting that the accuracy of the commercial-tool-flow-based flows is generally lower than that of the OpenROAD-based flow.  The reason why OpenROAD provides better accuracy is because we have visibility into the full $\pi$-model, while the commercial tool API only provides a view of the total $C_{\mbox{\scriptsize load}}$.  However, the error in the commercial flow is still acceptable. To examine this further, we show histograms of the distribution of wirelengths for a representative design, swerv\_wrapper in 45nm, under both flows. Fig.~\ref{fig:wirelength_distribution} shows the number of nets with lengths greater than 100$\mu$m under both flows.  From the histograms, it is apparent that the designs generated by the commercial flow have fewer long wires as shown in Fig.~\ref{fig:wirelength_distribution}; this can be attributed to lower utilizations for OpenROAD {\em vis a vis} the commercial tool flow. Due to these shorter wirelengths, the simpler $C_{\mbox{\scriptsize load}}$ feature is adequate to capture the wire capacitance, and is supplemented by the source-to-sink wirelength feature that acts as a surrogate for a wire resistance feature.  For the OpenROAD-based flow, with longer wires, we find that the use of these features results in significant accuracy loss, and that the $\pi$-model features are required to achieve the accuracy levels shown in Table~\ref{tbl:ml-sink-accuracy2}.}

\subsubsection{Path-level accuracy} We analyze the accuracy of the predictor in estimating the delays of a sample of paths in the circuit. These path delays are estimated by annotating the predicted parasitic values, wire delays, and wire slews into a timing engine. We compare the path slacks across the traditional flow and the ML-based flow.  \blueHL{Specifically, we analyze paths whose slack is below 40\% of the clock period, considering one worst-case path for each endpoint.}

\begin{figure}[tbp]
\begin{subfigure}{0.65\textwidth}
\includegraphics[width=\linewidth]{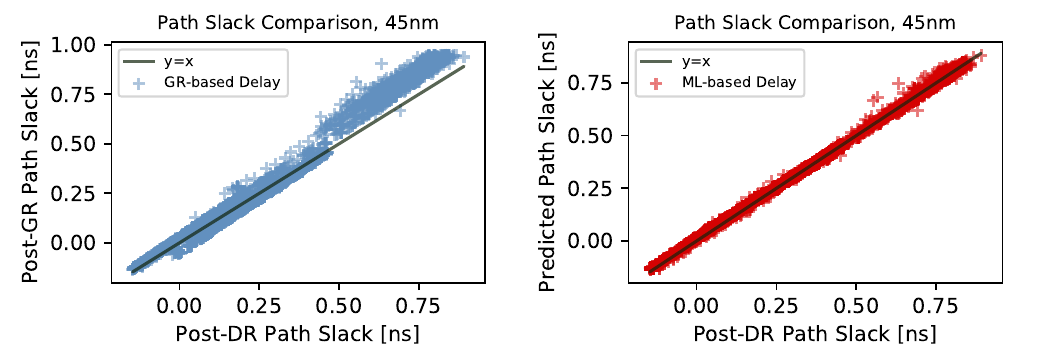}
\caption{ }\label{fig:path_slack_45nm_OR}
\end{subfigure}
\begin{subfigure}{0.65\textwidth}
\includegraphics[width=\linewidth]{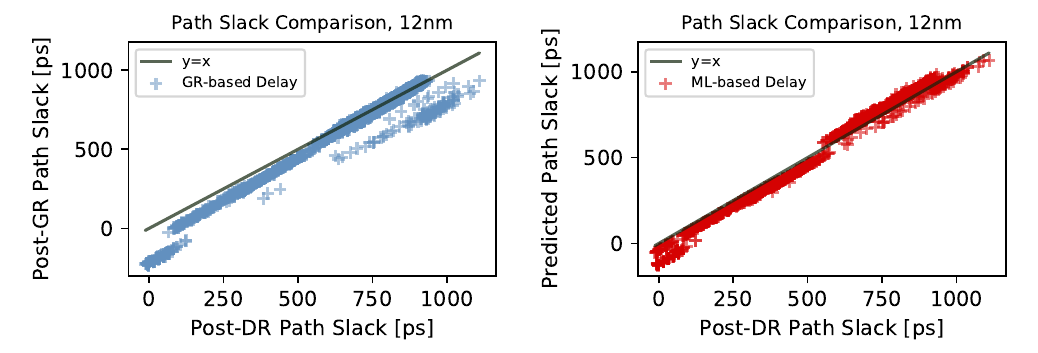}
\caption{ }\label{fig:path_slack_12nm_OR}
\end{subfigure}
\caption{Path slack comparison for (a) swerv\_wrapper 45nm and (b) coyote 12nm.}
\label{fig:path_slack_OR}
\end{figure}

\blueHL{Fig.~\ref{fig:path_slack_OR} shows an example of the slack comparison for (a) swerv\_wrapper in 45nm technology and (b) coyote in 12nm technology. The figures on the left show the discrepancy in path slack between GR and DR timing estimates and the figures on the right show the ML-corrected path slacks versus the post-DR path slacks. With the ML-based timing correction applied after GR, the post-GR path slacks have a better match with post-DR slacks.}

\blueHL{The last six columns in Table~\ref{tbl:ml-sink-accuracy2} compare the mean, maximum and standard deviation of path delay \%errors from the ML model against the \%errors from the traditional flow for multiple designs in both OpenROAD and the commercial tool flow. The mean path delay \%error is defined as the mean of the absolute percentage difference between the ML-based delays and the post-DR path delays, using the clock constraint for each design as the (normalization) reference. The traditional flow has a higher mean, maximum, and standard deviation of \%error when compared to the ML-based flow, with just one exception -- the standard deviation of bp\_be 45nm in OpenROAD is slightly worse than with the traditional flow. 
% \redHL{Noting that for a fixed timing specification, the \%error in path delay is the same as the \%error in path slack, this}
This indicates that on average the ML-based post-GR delays correlate better with post-DR delays than the traditional-flow-based delays. The path delay accuracy is improved by our approach, with the average error reducing from 2.50\% to 1.11\%.
This enables timing optimizations to buffer nets and resize logic gates on truly critical paths.}

\subsection{Impact on post-DR outcomes}

\begin{table}
\centering
\caption{\blueHL{Impact of ML-based prediction for designs with macros on post-DR metrics for traditional, ground-truth and ML-based flows in OpenROAD.}}
\label{tbl:ml-impact-on-DR-OR}
\resizebox{\linewidth}{!}{\begin{tabular}{|c|c|c|c|c|c|c|c|c|c|c|c|c|c|c|} 
\hline
\multirow{2}{*}{\textbf{Design}}                                                   & \multirow{2}{*}{\textbf{Tech}} & \multirow{2}{*}{\begin{tabular}[c]{@{}c@{}}\textbf{Die size}\\\textbf{(mm\textsuperscript{2})}\end{tabular}} & \multirow{2}{*}{\greenHL{\textbf{Utilization}}} & \multicolumn{4}{c|}{\textbf{Post-DR WS (ns)}}                                                                                                                                & \multicolumn{4}{c|}{\textbf{Post-DR TNS (ns)}}                                                                                                                                                                       & \multicolumn{3}{c|}{\textbf{Runtimes (s)}}                                                                                                                      \\ 
\cline{5-15}
                                                                                   &                                &                                                                                                              &                                       & \textbf{Trad.} & \begin{tabular}[c]{@{}c@{}}\textbf{~ML}\\\textbf{based}\end{tabular} & \textbf{\cite{Chhabria22}}  & \begin{tabular}[c]{@{}c@{}}\textbf{Ground}\\\textbf{Truth}\end{tabular} & \textbf{Trad.} & \begin{tabular}[c]{@{}c@{}}\textbf{ML}\\\textbf{based}\end{tabular} & \begin{tabular}[c]{@{}c@{}}\textbf{\cite{Chhabria22}}\\\end{tabular} & \begin{tabular}[c]{@{}c@{}}\textbf{Ground}\\\textbf{Truth}\end{tabular} & \textbf{Trad.} & \begin{tabular}[c]{@{}c@{}}\textbf{ML}\\\textbf{based}\end{tabular} & \begin{tabular}[c]{@{}c@{}}\textbf{Ground}\\\textbf{Truth}\end{tabular}  \\ 
\hline
\multirow{3}{*}{\begin{tabular}[c]{@{}c@{}}swerv\_wrapper\\CLK=2.5ns\end{tabular}} & \multirow{9}{*}{45nm}          & 1.20x1.09                                                                                                    & 45.87\%                               & -0.14          & -0.14                                                                & -0.13      & -0.13                                                                   & -23.45         & \textcolor[rgb]{0,0.502,0}{\textbf{-21.96}}                         & -22.78                                              & -22.58                                                                  & 1629           & 1657                                                                & 1995                                                                     \\ 
\cline{3-15}
                                                                                   &                                & 1.36x1.20                                                                                                    & 39.22\%                               & -0.20          & -0.20                                                                & -0.20      & -0.20                                                                   & -38.40         & \textcolor[rgb]{0,0.502,0}{-37.22}                                  & -37.69                                              & -35.03                                                                  & 1601           & 1630                                                                & 1881                                                                     \\ 
\cline{3-15}
                                                                                   &                                & 1.55x1.34                                                                                                    & 30.71\%                               & -0.24          & \textcolor[rgb]{0,0.502,0}{-0.23}                                    & -0.23      & -0.21                                                                   & -57.22         & \textcolor[rgb]{0,0.502,0}{\textbf{-43.22}}                         & -50.16                                              & -49.16                                                                  & 1588           & 1613                                                                & 1898                                                                     \\ 
\cline{1-1}\cline{3-15}
\multirow{3}{*}{\begin{tabular}[c]{@{}c@{}}bp\_fe\\CLK=2.2ns\end{tabular}}         &                                & 0.78x0.63                                                                                                    & 43.73\%                               & -0.15          & \textcolor[rgb]{0,0.502,0}{-0.10}                                    & -0.13      & -0.10                                                                   & -1.60          & \textcolor[rgb]{0,0.502,0}{-0.95}                                   & -1.40                                               & -0.72                                                                   & 700            & 707                                                                 & 891                                                                      \\ 
\cline{3-15}
                                                                                   &                                & 0.84x0.68                                                                                                    & 37.45\%                               & 0.01           & \textcolor[rgb]{0,0.502,0}{0.02}                                     & 0.01       & 0.02                                                                    & 0.00           & 0.00                                                                & 0.00                                                & 0.00                                                                    & 619            & 628                                                                 & 681                                                                      \\ 
\cline{3-15}
                                                                                   &                                & 0.99x0.79                                                                                                    & 27.09\%                               & -0.01          & \textcolor[rgb]{0,0.502,0}{\textbf{0.02}}                            & -0.01      & 0.00                                                                    & -0.01          & \textcolor[rgb]{0,0.502,0}{0.00}                                    & -0.03                                               & 0.00                                                                    & 598            & 604                                                                 & 709                                                                      \\ 
\cline{1-1}\cline{3-15}
\multirow{3}{*}{\begin{tabular}[c]{@{}c@{}}bp\_be\\CLK=2.8ns\end{tabular}}         &                                & 0.75x0.75                                                                                                    & 41.82\%                               & -0.18          & \textcolor[rgb]{0,0.502,0}{-0.17}                                    & -0.17      & -0.15                                                                   & -9.63          & \textcolor[rgb]{0,0.502,0}{-7.82}                                   & -7.92                                               & -6.60                                                                   & 792            & 798                                                                 & 1217                                                                     \\ 
\cline{3-15}
                                                                                   &                                & 0.79x0.78                                                                                                    & 38.30\%                               & -0.15          & \textcolor[rgb]{0,0.502,0}{-0.12}                                    & -0.12      & -0.10                                                                   & -10.86         & \textcolor[rgb]{0,0.502,0}{-6.81}                                   & -7.76                                               & -4.15                                                                   & 768            & 774                                                                 & 988                                                                      \\ 
\cline{3-15}
                                                                                   &                                & 0.98x0.92                                                                                                    & 25.95\%                               & -0.09          & \textcolor[rgb]{0,0.502,0}{\textbf{-0.07}}                           & -0.09      & -0.10                                                                   & -2.30          & \textcolor[rgb]{0,0.502,0}{-2.10}                                   & -2.63                                               & -2.09                                                                   & 622            & 629                                                                 & 889                                                                      \\ 
\hline
\multirow{3}{*}{\begin{tabular}[c]{@{}c@{}}swerv\_wrapper\\CLK=1.2ns\end{tabular}} & \multirow{6}{*}{12nm}          & 0.64x0.48                                                                                                    & 49.08\%                               & -0.48          & \textcolor[rgb]{0,0.502,0}{\textbf{-0.24}}                           & -0.32      & -0.36                                                                   & -207.68        & \textcolor[rgb]{0,0.502,0}{-202.39}                                 & -204.37                                             & -200.55                                                                 & 6041           & 6060                                                                & 13861                                                                    \\ 
\cline{3-15}
                                                                                   &                                & 0.70x0.54                                                                                                    & 39.86\%                               & -0.35          & \textcolor[rgb]{0,0.502,0}{\textbf{-0.23}}                           & -0.27      & -0.33                                                                   & -364.02        & \textcolor[rgb]{0,0.502,0}{-343.81}                                 & -347.05                                             & -337.56                                                                 & 5418           & 5433                                                                & 12072                                                                    \\ 
\cline{3-15}
                                                                                   &                                & 0.80x0.64                                                                                                    & 29.39\%                               & -0.21          & \textcolor[rgb]{0,0.502,0}{-0.18}                                    & -0.20      & -0.18                                                                   & -306.18        & \textcolor[rgb]{0,0.502,0}{\textbf{-234.98}}                        & -284.44                                             & -237.94                                                                 & 4846           & 4865                                                                & 10900                                                                    \\ 
\cline{1-1}\cline{3-15}
\multirow{3}{*}{\begin{tabular}[c]{@{}c@{}}coyote\\CLK=3.2ns\end{tabular}}         &                                & 0.66x0.66                                                                                                    & 49.08\%                               & -0.02          & \textcolor[rgb]{0,0.502,0}{0.05}                                     & 0.04       & 0.09                                                                    & -0.14          & \textcolor[rgb]{0,0.502,0}{0.00}                                    & 0.00                                                & 0.00                                                                    & 7243           & 7251                                                                & 13405                                                                    \\ 
\cline{3-15}
                                                                                   &                                & 0.75x0.75                                                                                                    & 35.87\%                               & -0.27          & \textcolor[rgb]{0,0.502,0}{-0.15}                                    & -0.19      & -0.14                                                                   & -24.73         & \textcolor[rgb]{0,0.502,0}{\textbf{-12.93}}                         & -20.50                                              & -14.06                                                                  & 6029           & 6038                                                                & 10883                                                                    \\ 
\cline{3-15}
                                                                                   &                                & 0.85x0.85                                                                                                    & 27.91\%                               & -0.08          & \textcolor[rgb]{0,0.502,0}{-0.05}                                    & -0.06      & -0.04                                                                   & -1.21          & \textcolor[rgb]{0,0.502,0}{-0.94}                                   & -1.04                                               & -0.75                                                                   & 6054           & 6064                                                                & 11465                                                                    \\
\hline
\end{tabular}}
\end{table}

\begin{table}
\centering
\caption{\blueHL{Impact of ML-based prediction on post-DR metrics for traditional, ground-truth and ML-based flows in a commercial tool.}}
\label{tbl:ml-impact-on-DR-comm}
\resizebox{0.8\linewidth}{!}{\begin{tabular}{|c|c|c|c|c|c|c|c|c|c|c|c|c|} 
\hline
\multirow{2}{*}{\textbf{Design}}                                                   & \multirow{2}{*}{\textbf{Tech}} & \multirow{2}{*}{\begin{tabular}[c]{@{}c@{}}\textbf{Die size}\\\textbf{(mm\textsuperscript{2})}\end{tabular}} & \multirow{2}{*}{\greenHL{\textbf{Utilization}}} & \multicolumn{3}{c|}{\textbf{Post-DR WS (ns)}}                                                                                                                   & \multicolumn{3}{c|}{\textbf{Post-DR TNS (ns)}}                                                                                                                 & \multicolumn{3}{c|}{\textbf{Runtime(s)}}                                                                                                                        \\ 
\cline{5-13}
                                                                                   &                                &                                                                                                              &                                       & \textbf{Trad.} & \begin{tabular}[c]{@{}c@{}}\textbf{~ML}\\\textbf{based}\end{tabular} & \begin{tabular}[c]{@{}c@{}}\textbf{Ground}\\\textbf{Truth}\end{tabular} & \textbf{Trad.} & \begin{tabular}[c]{@{}c@{}}\textbf{ML}\\\textbf{based}\end{tabular} & \begin{tabular}[c]{@{}c@{}}\textbf{Ground}\\\textbf{Truth}\end{tabular} & \textbf{Trad.} & \begin{tabular}[c]{@{}c@{}}\textbf{ML}\\\textbf{based}\end{tabular} & \begin{tabular}[c]{@{}c@{}}\textbf{Ground}\\\textbf{Truth}\end{tabular}  \\ 
\hline
\multirow{3}{*}{\begin{tabular}[c]{@{}c@{}}swerv\_wrapper\\CLK=2.2ns\end{tabular}} & \multirow{9}{*}{45nm}          & 1.18x1.18                                                                                                    & 48.85\%                               & -0.04          & \textcolor[rgb]{0,0.502,0}{-0.02}                                    & -0.01                                                                   & -0.22          & \textcolor[rgb]{0,0.502,0}{-0.12}                                   & -0.01                                                                   & 629            & 661                                                                 & 2435                                                                     \\ 
\cline{3-13}
                                                                                   &                                & 1.25x1.25                                                                                                    & 40.76\%                               & 0.00           & 0.00                                                                 & 0.00                                                                    & 0.00           & 0.00                                                                & 0.00                                                                    & 501            & 533                                                                 & 1995                                                                     \\ 
\cline{3-13}
                                                                                   &                                & 1.44x1.44                                                                                                    & 36.91\%                               & -0.08          & -0.08                                                                & -0.04                                                                   & -0.89          & \textcolor[rgb]{0,0.502,0}{-0.52}                                   & -0.40                                                                   & 507            & 602                                                                 & 2243                                                                     \\ 
\cline{1-1}\cline{3-13}
\multirow{3}{*}{\begin{tabular}[c]{@{}c@{}}bp\_fe\\CLK=1.6ns\end{tabular}}         &                                & 0.63x0.63                                                                                                    & 50.72\%                               & -0.29          & \textcolor[rgb]{0,0.502,0}{-0.24}                                    & -0.01                                                                   & -10.96         & \textcolor[rgb]{0,0.502,0}{-8.75}                                   & 0.00                                                                    & 257            & 265                                                                 & 867                                                                      \\ 
\cline{3-13}
                                                                                   &                                & 0.71x0.71                                                                                                    & 41.77\%                               & -0.31          & \textcolor[rgb]{0,0.502,0}{-0.15}                                    & -0.04                                                                   & -11.96         & \textcolor[rgb]{0,0.502,0}{-3.32}                                   & -0.10                                                                   & 317            & 325                                                                 & 1216                                                                     \\ 
\cline{3-13}
                                                                                   &                                & 0.82x0.82                                                                                                    & 30.76\%                               & -0.41          & \textcolor[rgb]{0,0.502,0}{-0.09}                                    & -0.03                                                                   & -14.46         & \textcolor[rgb]{0,0.502,0}{-1.20}                                   & -0.30                                                                   & 254            & 262                                                                 & 1009                                                                     \\ 
\cline{1-1}\cline{3-13}
\multirow{3}{*}{\begin{tabular}[c]{@{}c@{}}bp\_be\\CLK=2.0ns\end{tabular}}         &                                & 0.66x0.66                                                                                                    & 53.26\%                               & -0.04          & -0.04                                                                & -0.01                                                                   & -2.65          & \textcolor[rgb]{0,0.502,0}{-1.20}                                   & -0.04                                                                   & 401            & 415                                                                 & 1659                                                                     \\ 
\cline{3-13}
                                                                                   &                                & 0.74x0.74                                                                                                    & 42.83\%                               & -0.04          & -0.04                                                                & -0.01                                                                   & -2.35          & \textcolor[rgb]{0,0.502,0}{-1.01}                                   & -0.03                                                                   & 342            & 356                                                                 & 1445                                                                     \\ 
\cline{3-13}
                                                                                   &                                & 0.85x0.85                                                                                                    & 32.25\%                               & -0.03          & \textcolor[rgb]{0,0.502,0}{-0.01}                                    & -0.01                                                                   & -1.49          & \textcolor[rgb]{0,0.502,0}{-0.06}                                   & -0.03                                                                   & 336            & 351                                                                 & 1351                                                                     \\ 
\hline
\multirow{3}{*}{\begin{tabular}[c]{@{}c@{}}swerv\_wrapper\\CLK=1.0ns\end{tabular}} & \multirow{6}{*}{12nm}          & 0.52x0.52                                                                                                    & 46.85\%                               & -0.11          & \textcolor[rgb]{0,0.502,0}{-0.08}                                    & 0.01                                                                    & -18.68         & \textcolor[rgb]{0,0.502,0}{-5.28}                                   & 0.00                                                                    & 2320           & 2346                                                                & 9291                                                                     \\ 
\cline{3-13}
                                                                                   &                                & 0.58x0.58                                                                                                    & 38.14\%                               & 0.18           & 0.10                                                                 & 0.29                                                                    & 0.00           & 0.00                                                                & 0.00                                                                    & 2126           & 2152                                                                & 8498                                                                     \\ 
\cline{3-13}
                                                                                   &                                & 0.67x0.67                                                                                                    & 30.10\%                               & -0.28          & \textcolor[rgb]{0,0.502,0}{\textbf{-0.15}}                           & -0.18                                                                   & -53.91         & \textcolor[rgb]{0,0.502,0}{-49.30}                                  & -47.37                                                                  & 2897           & 2923                                                                & 9775                                                                     \\ 
\cline{1-1}\cline{3-13}
\multirow{3}{*}{\begin{tabular}[c]{@{}c@{}}coyote\\CLK=3.2ns\end{tabular}}         &                                & 0.56x0.56                                                                                                    & 47.87\%                               & -0.09          & \textcolor[rgb]{0,0.502,0}{-0.01}                                    & 0.01                                                                    & -0.80          & \textcolor[rgb]{0,0.502,0}{-0.01}                                   & 0.01                                                                    & 3455           & 3514                                                                & 13804                                                                    \\ 
\cline{3-13}
                                                                                   &                                & 0.59x0.59                                                                                                    & 42.64\%                               & -0.15          & \textcolor[rgb]{0,0.502,0}{-0.05}                                    & -0.02                                                                   & -2.87          & \textcolor[rgb]{0,0.502,0}{-0.12}                                   & -0.02                                                                   & 3986           & 4045                                                                & 15872                                                                    \\ 
\cline{3-13}
                                                                                   &                                & 0.68x0.68                                                                                                    & 38.33\%                               & -0.14          & \textcolor[rgb]{0,0.502,0}{-0.09}                                    & 0.00                                                                    & -1.07          & \textcolor[rgb]{0,0.502,0}{-0.48}                                   & 0.00                                                                    & 3698           & 3757                                                                & 14715                                                                    \\
\hline
\end{tabular}}
\end{table}

\begin{figure}[htbp]

\centerline{\includegraphics[width=0.7\linewidth]{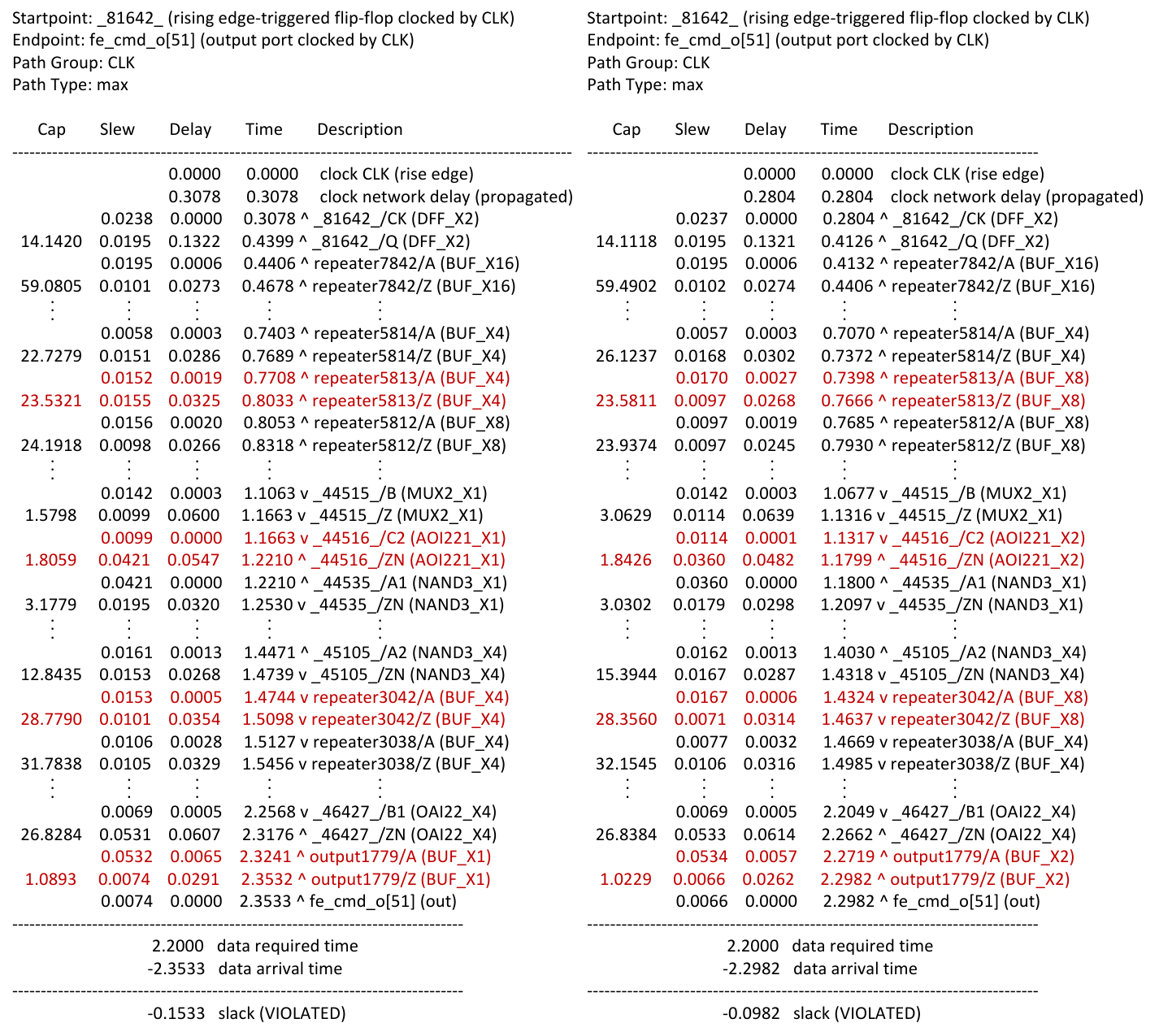}}
\caption{A critical path from bp\_be 45nm after DR, from the traditional flow (left) and from the ML-based flow (right).}
\label{fig:timing-path-comparisons}
\end{figure}

The three ML models are applied to the timing analysis step before post-GR timing optimization transforms, \blueHL{based on sizing and buffering}, in the physical design flow. \blueHL{We compare our post-DR outcomes against the flows in Fig.~\ref{fig:flows}(a) and (b).
Tables~\ref{tbl:ml-impact-on-DR-OR} and~\ref{tbl:ml-impact-on-DR-comm} compare post-DR WS and post-DR TNS for the \greenHL{four flows} applied in OpenROAD and in a commercial tool. The ground-truth flow optimizes paths that are critical as per the DR-based parasitics, while the traditional flow optimizes paths that may or may not be critical post-DR. For example,
Fig.~\ref{fig:timing-path-comparisons} shows a timing path from OpenSTA after DR. At left, we see the post-DR timing path from the traditional flow, and at right, we list the same path from the ML-based flow. The post-GR timing optimization does not have an accurate estimation of the path that becomes critical after DR and hence does not have correct resizing or buffering, while our ML-based flow identifies this as a critical path during post-GR optimization, and resizes the gate and buffers the net to ensure that the path has better timing after DR. As a consequence of better optimization, we improve post-DR WS.}

\blueHL{Table~\ref{tbl:ml-impact-on-DR-OR} compares the post-DR WS and TNS from the ML-based flow and from the traditional flow, in OpenROAD. Out of 15 designs, the ML-based flow improves post-DR WS for 13 designs and improves post-DR TNS for 14 designs as highlighted in green, indicating effective timing optimization post-GR using the ML model. Of the remaining cases,  ``Trad." and ``ML'' achieve the same results. For the cases in bold green, the post-DR results from the ML-based flow are even better than the results from the ground-truth-based flow.  Based on a detailed analysis of the results, we attribute this to two reasons. First, the path slack from the ML-based estimation can be pessimistic due to prediction error; this makes the tool flow optimize delays more aggressively during post-GR optimization and thus helps improve post-DR WNS. Second, the routing of the worst post-DR timing path of the ground-truth-based flow has larger detours than the routing of the same path in the ML-based flow, which leads to more parasitics and worse timing, even though the two flows achieve the same buffering and sizing during post-GR optimization.} 

Table~\ref{tbl:ml-impact-on-DR-OR} also compares the ML-based flow to the flow proposed in the preliminary version of this paper in~\cite{Chhabria22}. We find that the post-DR WS of 10 designs and the post-DR TNS of 13 designs are improved by our introduction of new features.

\begin{figure}[tbp]
\begin{subfigure}{0.45\textwidth}
\includegraphics[width=\linewidth]{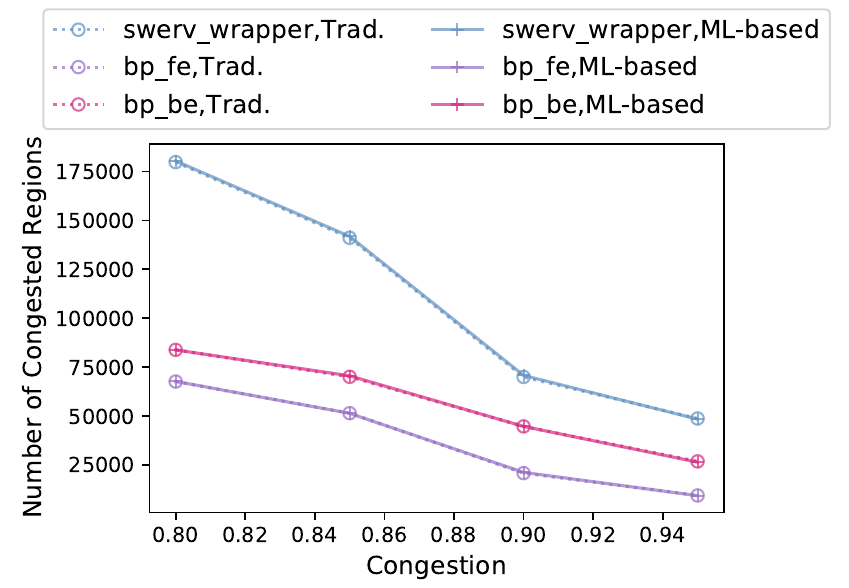}
\caption{ }\label{fig:congestion_45nm_OR}
\end{subfigure}
\begin{subfigure}{0.45\textwidth}
\includegraphics[width=\linewidth]{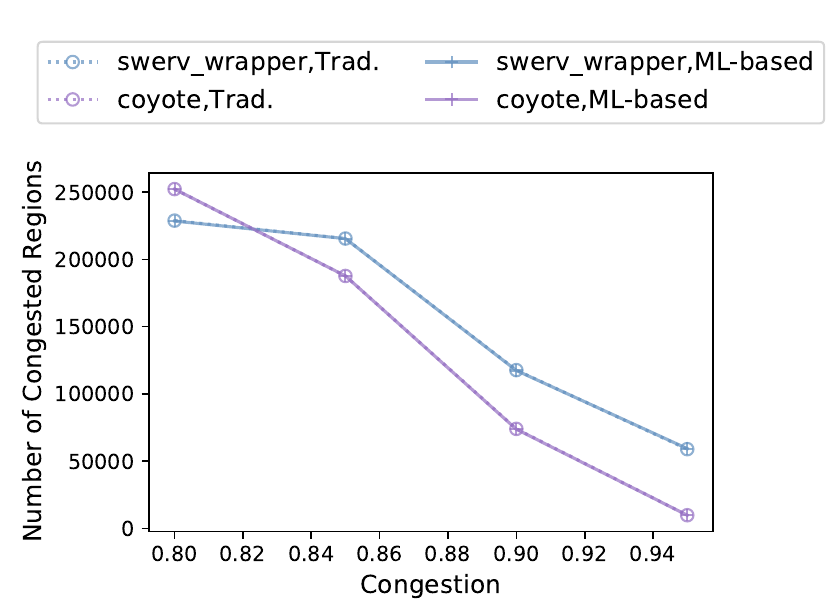}
\caption{ }\label{fig:congestion_12nm}
\end{subfigure}
\caption{Number of post-DR congested regions in the traditional and ML-based flows in (a) 45nm designs and (b) 12nm designs.}
\label{fig:num-congestion-gcells}
\end{figure}

To determine the impact of ML-based post-GR timing optimizations on routability, we analyze congestion under traditional and ML-based flows. We define a GCell to be congested if its congestion exceeds a specified threshold. Fig.~\ref{fig:num-congestion-gcells} shows the number of congested GCells in the traditional and ML-based flows, for different thresholds. The ML-based model does not increase the number of congested GCells (i.e., the traces are superposed), indicating that it does not impact routability. 

\blueHL{The last three columns of Table \ref{tbl:ml-impact-on-DR-OR} compare the runtimes of three different flows. We see that the ML-based flow can leverage prediction of post-DR timing at the cost of a few tens of seconds, without time-intensive DR. 
The ML-based flow achieves comparable solutions to the ground-truth-based flow, with an average speedup of 31.87\% over the ground-truth flow. When compared to the traditional flow, 
the ML-based flow is only marginally (0.84\%) slower than the traditional flow, which is acceptable as the model can achieve a quality that is closer to, and sometimes better than, the ground-truth flow.}

\blueHL{Table~\ref{tbl:ml-impact-on-DR-comm} compares the post-DR WS and TNS results from the commercial tool flow. Out of 15 designs, the ML-based flow improves post-DR WS for 10 designs and improves post-DR TNS 13 designs. Four designs have the same post-DR WS in the traditional flow and the ML-based flow. One design (swerv\_wrapper in 12nm) improves the WS of 0.18ns from the traditional flow to 0.10ns using our approach, with lower buffering and sizing cost. For post-DR TNS, the ML-based flow shows better results in 13 designs, and the remaining designs do not have path violations.}

\subsection{Generalization to different clock periods}

\begin{figure}[tbp]
\includegraphics[width=\linewidth]{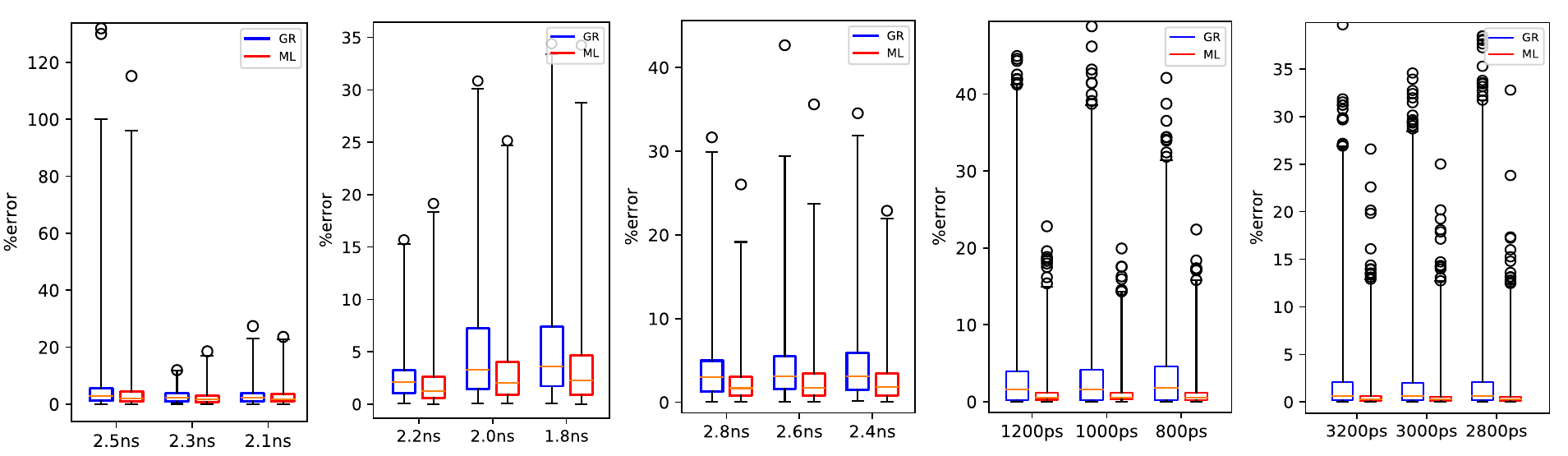}
\caption{\blueHL{Sensitivity of wire delay models to different clock periods for (from left to right) swerv\_wrapper, bp\_fe, bp\_be in 45nm, and swerv\_wrapper, coyote in 12nm.}}
\label{fig:CLK-sensitivity}
\end{figure}

\blueHL{In the experiments above, the designs used for training have been implemented at a single target clock period. We now analyze the ability of our ML models to generalize with respect to unseen designs generated by using different clock periods from the one used for training. To test the sensitivity of the models to different clock periods, we implement each design with two more clock periods and extract the input features for running ML inferences on two unseen designs. Fig.~\ref{fig:CLK-sensitivity} shows the prediction accuracy for all designs generated using three different clock constraints. 
\greenHL{Here, the model presented in Tables~\ref{tbl:ml-sink-accuracy2} and \ref{tbl:ml-impact-on-DR-OR} is used for the leftmost clock period.} The leftmost clock period of each plot is used for training, which is also denoted in blue in Table~\ref{tbl:clk-sensitivity}, and the other two clock periods are unseen during the training. The mean \%error for each case is listed in Table~\ref{tbl:clk-sensitivity}. In the box-whisker plot, the boxes indicate the 25$^{\rm th}$ percentile, 50$^{\rm th}$ percentile (the median, shown by a yellow line), and 75$^{\rm th}$ percentile of the prediction error of the traditional method in GR and ML-based method. For the ML-based method, the box and the median are generally lower than for the traditional method, indicating that ML models have good ability to generalize with respect to unseen designs with different clock constraints. 
%Although the median slew error for the ML-based method for swerv\_wrapper under a 2.3ns clock constraint, both the mean \%error from ML-based method and from traditional methods are 0.06\% according to Table~\ref{tbl:clk-sensitivity}. 
The table summarizes the mean \%error of the prediction for different clock periods, one of which used for training and is highlighted in blue. The smaller ML prediction errors for designs in 12nm also indicate that our models generalize well in 12nm.}

\begin{table}
\centering
\caption{\blueHL{Mean \%error of prediction for designs generated using different clock periods.}}
\label{tbl:clk-sensitivity}
\resizebox{0.8\linewidth}{!}{\begin{tabular}{|c|c|c|c|c|c|c|} 
\hline
\textbf{Design}                 & \textbf{Tech}         & \begin{tabular}[c]{@{}c@{}}\textbf{Clock}\\\textbf{period}\end{tabular} & \textbf{ML wire delay} & \textbf{GR wire delay} & \textbf{ML wire slew} & \textbf{GR wire slew}  \\ 
\hline
\multirow{3}{*}{swerv\_wrapper} & \multirow{9}{*}{45nm} & \textcolor{blue}{2.50ns}                                                & 0.84\%                 & 4.27\%                 & 0.04\%                & 0.20\%                 \\ 
\cline{3-7}
                                &                       & 2.30ns                                                                  & 2.40\%                 & 4.50\%                 & 0.06\%                & 0.06\%                 \\ 
\cline{3-7}
                                &                       & 2.10ns                                                                  & 3.35\%                 & 5.61\%                 & 0.07\%                & 0.09\%                 \\ 
\cline{1-1}\cline{3-7}
\multirow{3}{*}{bp\_fe}         &                       & \textcolor{blue}{2.20ns}                                                & 0.64\%                 & 3.63\%                 & 0.04\%                & 0.24\%                 \\ 
\cline{3-7}
                                &                       & 2.00ns                                                                  & 2.91\%                 & 3.15\%                 & 0.23\%                & 0.31\%                 \\ 
\cline{3-7}
                                &                       & 1.80ns                                                                  & 2.40\%                 & 3.78\%                 & 0.16\%                & 0.25\%                 \\ 
\cline{1-1}\cline{3-7}
\multirow{3}{*}{bp\_be}         &                       & \textcolor{blue}{2.80ns}                                                & 2.16\%                 & 3.00\%                 & 0.07\%                & 0.12\%                 \\ 
\cline{3-7}
                                &                       & 2.60ns                                                                  & 2.33\%                 & 3.13\%                 & 0.12\%                & 0.13\%                 \\ 
\cline{3-7}
                                &                       & 2.40ns                                                                  & 2.51\%                 & 3.70\%                 & 0.11\%                & 0.13\%                 \\ 
\hline
\multirow{3}{*}{swerv\_wrapper} & \multirow{6}{*}{12nm} & \textcolor{blue}{1.20ns}                                                & 0.87\%                 & 4.26\%                 & 0.04\%                & 0.70\%                 \\ 
\cline{3-7}
                                &                       & 1.00ns                                                                  & 1.21\%                 & 4.03\%                 & 0.12\%                & 0.56\%                 \\ 
\cline{3-7}
                                &                       & 0.80ns                                                                  & 1.22\%                 & 4.79\%                 & 0.26\%                & 1.31\%                 \\ 
\cline{1-1}\cline{3-7}
\multirow{3}{*}{coyote}         &                       & \textcolor{blue}{3.20ns}                                                & 0.60\%                 & 3.47\%                 & 0.22\%                & 1.97\%                 \\ 
\cline{3-7}
                                &                       & 3.00ns                                                                  & 0.75\%                 & 3.23\%                 & 0.32\%                & 1.99\%                 \\ 
\cline{3-7}
                                &                       & 2.80ns                                                                  & 0.81\%                 & 3.17\%                 & 0.31\%                & 1.68\%                 \\
\hline
\end{tabular}}
\end{table}

\subsection{Noise impact on model performance}

\begin{table}
\centering
\caption{\blueHL{Impact of training-stage noise, for various values of the noise standard deviation, on prediction accuracy.}}
\label{tbl:noise-accuracy}
\resizebox{\linewidth}{!}{
\begin{tabular}{|c|c|c|c|c|c|c|c|c|} 
\hline
\multirow{2}{*}{\textbf{Design}}                                                    & \multirow{2}{*}{\textbf{Tech}} & \multirow{2}{*}{\textbf{Metrics}} & \multicolumn{3}{c|}{\textbf{Wire Delay}}                                                             & \multicolumn{3}{c|}{\textbf{Wire Slew}}                                                             \\ 
\cline{4-9}
                                                                                    &                                &                                   & Trad.~   & ML~      & ML-Noise(0.01, 0.05, 0.1)                                                      & Trad.~  & ML      & ML-Noise(0.01, 0.05, 0.1)                                                       \\ 
\hline
\multirow{3}{*}{\begin{tabular}[c]{@{}c@{}}swerv\_wrapper\\clk=2.5ns\end{tabular}}  & \multirow{9}{*}{45nm}          & Mean \%error                      & 4.27\%   & 0.84\%   & (1.37\%, 1.89\%, 2.65\%)                                                       & 0.20\%  & 0.04\%  & (0.09\%, 0.20\%, \textcolor{red}{0.33\%})                                       \\ 
\cline{3-9}
                                                                                    &                                & Max \%error                       & 149.41\% & 115.19\% & (124.90\%, 133.41\%, 137.82\%)                                                 & 40.63\% & 19.81\% & (33.79\%, 33.76\%, 34.38\%)                                                     \\ 
\cline{3-9}
                                                                                    &                                & Std. dev. \%error                 & 7.28\%   & 2.78\%   & (4.34\%, 5.24\%, 5.73\%)                                                       & 1.22\%  & 0.34\%  & (0.65\%, 0.76\%, 0.75\%)                                                        \\ 
\cline{1-1}\cline{3-9}
\multirow{3}{*}{\begin{tabular}[c]{@{}c@{}}bp\_fe\\clk=2.2ns\end{tabular}}          &                                & Mean \%error                      & 3.63\%   & 0.64\%   & (1.60\%,\textcolor{red}{ 4.39\%}, \textcolor{red}{7.06\%})                     & 0.24\%  & 0.04\%  & (0.10\%, 0.31\%, 0.51\%)                                                        \\ 
\cline{3-9}
                                                                                    &                                & Max \%error                       & 15.68\%  & 19.14\%  & (\textcolor{red}{25.32\%}, \textcolor{red}{27.98\%}, \textcolor{red}{47.65\%}) & 5.83\%  & 4.42\%  & (3.70\%, 5.36\%, \textcolor{red}{7.39\%})                                       \\ 
\cline{3-9}
                                                                                    &                                & Std. dev. \%error                 & 4.65\%   & 0.91\%   & (1.47\%, 4.17\%, \textcolor{red}{7.31\%})                                      & 0.98\%  & 0.13\%  & (0.22\%, 0.52\%, 0.92\%)                                                        \\ 
\cline{1-1}\cline{3-9}
\multirow{3}{*}{\begin{tabular}[c]{@{}c@{}}bp\_be\\clk=2.8ns\end{tabular}}          &                                & Mean \%error                      & 3.00\%   & 2.16\%   & (2.64\%, 2.71\%, \textcolor{red}{4.36\%})                                      & 0.12\%  & 0.07\%  & (0.08\%, \textcolor{red}{0.19\%}, \textcolor{red}{0.33\%})                      \\ 
\cline{3-9}
                                                                                    &                                & Max \%error                       & 55.64\%  & 34.58\%  & (31.64\%, 39.62\%, 47.77\%)                                                    & 10.45\% & 9.15\%  & (\textcolor{red}{14.62\%}, \textcolor{red}{18.69\%}, \textcolor{red}{28.10\%})  \\ 
\cline{3-9}
                                                                                    &                                & Std. dev. \%error                 & 3.88\%   & 3.42\%   & (3.78\%, 3.08\%, 3.23\%)                                                       & 0.38\%  & 0.25\%  & (\textcolor{red}{0.44\%}, \textcolor{red}{0.62\%}, \textcolor{red}{0.91\%})     \\ 
\hline
\multirow{3}{*}{\begin{tabular}[c]{@{}c@{}}swerv\_wrapper\\clk=1200ps\end{tabular}} & \multirow{6}{*}{12nm}          & Mean \%error                      & 4.26\%   & 0.87\%   & (1.03\%, 2.73\%, \textcolor{red}{4.89\%})                                      & 0.70\%  & 0.04\%  & (0.04\%, 0.09\%, 0.16\%)                                                        \\ 
\cline{3-9}
                                                                                    &                                & Max \%error                       & 44.95\%  & 22.80\%  & (25.90\%, 32.35\%, \textcolor{red}{67.92\%})                                   & 71.75\% & 29.81\% & (29.28\%, 29.65\%, 33.14\%)                                                     \\ 
\cline{3-9}
                                                                                    &                                & Std. dev. \%error                 & 5.76\%   & 0.93\%   & (0.99\%, 2.69\%, 5.25\%)                                                       & 3.85\%  & 0.17\%  & (0.18\%, 0.22\%, 0.32\%)                                                        \\ 
\cline{1-1}\cline{3-9}
\multirow{3}{*}{\begin{tabular}[c]{@{}c@{}}coyote\\clk=3200ps\end{tabular}}         &                                & Mean \%error                      & 3.47\%   & 0.60\%   & (0.68\%, 2.34\%, \textcolor{red}{4.16\%})                                      & 1.97\%  & 0.22\%  & (0.23\%, 0.80\%, 1.43\%)                                                        \\ 
\cline{3-9}
                                                                                    &                                & Max \%error                       & 39.65\%  & 26.58\%  & (27.65\%, 30.56\%, \textcolor{red}{63.40\%})                                   & 51.34\% & 40.13\% & (41.80\%, 43.21\%, 43.42\%)                                                     \\ 
\cline{3-9}
                                                                                    &                                & Std. dev. \%error                 & 4.73\%   & 0.88\%   & (0.91\%, 3.10\%, \textcolor{red}{5.87\%})                                      & 4.51\%  & 0.62\%  & (0.64\%, 1.10\%, 1.78\%)                                                        \\
\hline
\end{tabular}}
\end{table}
\begin{table}
\centering
\caption{\blueHL{Impact of noise with various standard deviation on post-DR results.}}
\label{tbl:noise-impact-on-DR}
\resizebox{\linewidth}{!}{
\begin{tabular}{|c|c|c|c|c|c|c|c|} 
\hline
\multirow{2}{*}{\textbf{Design}}                                   & \multirow{2}{*}{\textbf{Tech}} & \multicolumn{3}{c|}{\textbf{Post-DR WS (ns)}}                                        & \multicolumn{3}{c|}{\textbf{Post-DR TNS (ns)}}                                           \\ 
\cline{3-8}
                                                                   &                                & Trad. & ML-Noise(0, 0.01, 0.05, 0.1)                                   & Ground-truth & Trad.   & ML-Noise(0, 0.01, 0.05, 0.1)                                   & Ground-truth  \\ 
\hline
\begin{tabular}[c]{@{}c@{}}swerv\_wrapper\\clk=2.5ns\end{tabular}  & \multirow{3}{*}{45nm}          & -0.14 & (-0.14, -0.14, -0.14, \textcolor{red}{-0.15})                  & -0.13        & -23.45  & (-21.96, -22.58, \textcolor{red}{-23.52, -26.89})              & -22.58        \\ 
\cline{1-1}\cline{3-8}
\begin{tabular}[c]{@{}c@{}}bp\_fe\\clk=2.2ns\end{tabular}          &                                & -0.15 & (-0.10, -0.14, \textcolor{red}{-0.16}, \textcolor{red}{-0.21}) & -0.10        & -1.60   & (-0.95, -1.40, \textcolor{red}{-1.88}, \textcolor{red}{-2.30}) & -0.72         \\ 
\cline{1-1}\cline{3-8}
\begin{tabular}[c]{@{}c@{}}bp\_be\\clk=2.8ns\end{tabular}          &                                & -0.18 & (-0.17, -0.17, -0.18, -0.18)                                   & -0.15        & -9.63   & (-7.82, -7.65, -8.14, -8.77)                                   & -6.60         \\ 
\hline
\begin{tabular}[c]{@{}c@{}}swerv\_wrapper\\clk=1200ps\end{tabular} & \multirow{2}{*}{12nm}          & -0.48 & (-0.23, -0.28, -0.46, \textcolor{red}{-0.58})                  & -0.35        & -207.68 & (-202.38, -207.43,\textcolor{red}{ -209.80, -230.48})          & -200.55       \\ 
\cline{1-1}\cline{3-8}
\begin{tabular}[c]{@{}c@{}}coyote\\clk=3200ps\end{tabular}         &                                & -0.02 & (0.49, 0.47, 0.40, 0.34)                                       & 0.86         & -0.14   & (0, 0, 0, 0)                                                   & 0.00          \\
\hline
\end{tabular}}
\end{table}

\blueHL{As there is inherent noise from EDA tools, the data with noise collected from EDA tools and used in our ML-based flow can affect the DR solution quality. To study the impact of noise on the model accuracy and DR outcomes from the ML-based flow, we add Gaussian noise $\mathcal{N}(0,\sigma)$ to labels and train the models using the datasets with noise. A set of $\sigma$ values (0, 0.01, 0.05, 0.1) are selected to evaluate the robustness of the models to the noise with different standard deviations. We generate noise matrices that have the same dimension with our training dataset using three different random seeds for each $\sigma$ value, then add each noise matrix to the training dataset separately. We take the average value of the three predictions from the models trained by the three different training datasets as our prediction result. Then, we compare the results from a modified ML-based flow (where the ML models are replaced by the models trained with noisy data) to the results from the original ML-based flow.}

\blueHL{In Table~\ref{tbl:noise-accuracy}, the prediction errors of models trained with noisy data are compared to the traditional estimation and the prediction from models trained with a dataset without added noise. According to the results for $\sigma = 0.01$, ML models can handle the noise with $\sigma = 0.01$ value in most cases; the exceptions are maximum \%error for wire delay of bp\_fe in 45nm, and maximum and standard deviation of \%error for wire slew of bp\_be in 45nm. These are denoted in red, indicating no improvement compared to the estimation from traditional methods. According to the results for $\sigma = 0.05$, the ML models perform better on four out of five designs for both wire delay and wire delay. However, when $\sigma = 0.1$, which is approximately equal to the maximum wire delay and $2\times$ maximum slew in our datasets, the models trained by noisy data cannot make accurate predictions for most designs.}

\blueHL{The impact of noise on post-DR solution quality is summarized in Table~\ref{tbl:noise-impact-on-DR}. We compare the post-DR results from the ML-based flow using the models trained with noisy data, to the traditional flow and ground-truth-based flows. The ML-based flow generates better DR solutions than the traditional flow for all testcases when the noise has $\sigma = 0.01$. As we increase the $\sigma$ value, the DR results of some testcases denoted in red are no longer better than the results from the traditional flow, due to inaccurate timing prediction being used in post-GR timing optimization. For $\sigma = 0.05$ and $\sigma = 0.1$, only two designs out of five designs still see better solutions from the ML-based flow. }

\section{Conclusion}

We study an ML-based method to predict post-DR timing at the post-route stage.  Our models demonstrate better accuracy of timing and parasitic estimation at the post-GR stage compared to traditional methods, using both OpenROAD and a commercial tool flow. Our approach shows improvements in the DR solution quality; it is shown to be computationally efficient, and does not increase congestion compared to the prior methods. Moreover, our experimental results show that our models generalize well to designs generated using unseen clock periods, and that our flow generates better DR solutions even when models are trained using datasets with noise.

\begin{acks}
This work was supported in part by DARPA HR0011-18-2-0032 (The 
OpenROAD Project). The work of ABK is also supported in part by NSF CCF-2112665. 
\end{acks}

\bibliographystyle{ACM-Reference-Format}
\bibliography{references}

%\clearpage 

%\input coverletter

\end{document}